\documentclass[10pt,journal,compsoc]{IEEEtran}

\usepackage{graphicx}
\usepackage[section]{algorithm}
\usepackage{algorithmic}
\usepackage{amsmath}
\usepackage{bm}
\usepackage{ragged2e}
\usepackage[numbers,sort&compress]{natbib}
\usepackage{multirow}
\usepackage{longtable}
 \usepackage{rotating}
\renewcommand{\raggedright}{\leftskip=0pt \rightskip=0pt plus 0cm}

\begin{document}

\title{A Parallel Random Forest Algorithm for Big Data in a Spark Cloud Computing Environment}

\author{Jianguo~Chen,
        Kenli~Li,~\IEEEmembership{Senior Member, IEEE},
        Zhuo~Tang,~\IEEEmembership{Member, IEEE},\\
        Kashif~Bilal,
        Shui~Yu,~\IEEEmembership{Member, IEEE},
        Chuliang~Weng,~\IEEEmembership{Member, IEEE},
        and~Keqin~Li,~\IEEEmembership{Fellow, IEEE}% <-this % stops a space
\IEEEcompsocitemizethanks{\IEEEcompsocthanksitem Jianguo~Chen, Kenli~Li, Zhuo~Tang, and~Keqin~Li are with the College of Computer Science and Electronic Engineering, Hunan University, and the National Supercomputing Center in Changsha, Hunan, Changsha 410082, China.
\protect\\
Corresponding author: Kenli Li, Email: lkl@hnu.edu.cn.

\IEEEcompsocthanksitem Kashif Bilal is with the Qatar University,Doha 2713, Qatar, and the Comsats Institute of Information Technology, Islamabad 45550, Pakistan.

\IEEEcompsocthanksitem Shui Yu is with the School of Information Technology, Deakin University, Melbourne, Vic.3216, Australia.

\IEEEcompsocthanksitem Chuliang Weng is with the School of Computer Science and Software Engineering, Institute for Data Science and Engineering, East China Normal University, Shanghai 200241, China.

\IEEEcompsocthanksitem Keqin Li is also with the Department of Computer Science, State University of New York, New Paltz, NY 12561, USA.
}
}

\markboth{}%
{Shell \MakeLowercase{\textit{et al.}}: Bare Advanced Demo of IEEEtran.cls for Journals}

\IEEEtitleabstractindextext{%
\begin{abstract}
 \raggedright{
With the emergence of the big data age, the issue of how to obtain valuable knowledge from a dataset efficiently and accurately has attracted increasingly attention from both academia and industry.
This paper presents a Parallel Random Forest (PRF) algorithm for big data on the Apache Spark platform.
The PRF algorithm is optimized based on a hybrid approach combining data-parallel and task-parallel optimization.
From the perspective of data-parallel optimization, a vertical data-partitioning method is performed to reduce the data communication cost effectively, and a data-multiplexing method is performed is performed to allow the training dataset to be reused and diminish the volume of data.
From the perspective of task-parallel optimization, a dual parallel approach is carried out in the training process of RF, and a task Directed Acyclic Graph (DAG) is created according to the parallel training process of PRF and the dependence of the Resilient Distributed Datasets (RDD) objects.
Then, different task schedulers are invoked for the tasks in the DAG.
Moreover, to improve the algorithm's accuracy for large, high-dimensional, and noisy data, we perform a dimension-reduction approach in the training process and a weighted voting approach in the prediction process prior to parallelization.
Extensive experimental results indicate the superiority and notable advantages of the PRF algorithm over the relevant algorithms implemented by Spark MLlib and other studies in terms of the classification accuracy, performance, and scalability.
%With the expansion of the scale of the random forest model and the Spark cluster, the advantage of the PRF algorithm is more obvious.
}
\end{abstract}

}

\maketitle
\IEEEdisplaynontitleabstractindextext
\IEEEpeerreviewmaketitle

\section{Introduction}
\subsection{Motivation}
\IEEEPARstart{W}{ith} the continuous emergence of a variety of new information dissemination methods, and the rise of cloud computing and Internet of Things (IoT) technologies, data increase constantly with a high speed.
The scale of global data continuously increases at a rate of 2 times every two years \cite{ex1}.
The application value of data in every field is becoming more important than ever.
There exists a large amount of worthwhile information in available data.

The emergence of the big data age also poses serious problems and challenges besides the obvious benefits.
Because of business demands and competitive pressure, almost every business has a high demand for data processing in real-time and validity \cite{ex2}.
As a result, the first problem is how to mine valuable information from massive data efficiently and accurately.
At the same time, big data hold characteristics such as high dimensionality, complexity, and noise.
Enormous data often hold properties found in various input variables in hundreds or thousands of levels, while each one of them may contain a little information.
The second problem is to choose appropriate techniques that may lead to good classification performance for a high-dimensional dataset.
Considering the aforementioned facts, data mining and analysis for large-scale data have become a hot topic in academia and industrial research.

The speed of data mining and analysis for large-scale data has also attracted much attention from both academia and industry.
Studies on distributed and parallel data mining based on cloud computing platforms have achieved abundant favorable achievements \cite{ex3,ex4}.
Hadoop \cite{ex5} is a famous cloud platform widely used in data mining.
In \cite{ex6,ex7}, some machine learning algorithms were proposed based on the MapReduce model.
However, when these algorithms are implemented based on MapReduce, the intermediate results gained in each iteration are written to the Hadoop Distributed File System (HDFS) and loaded from it.
This costs much time for disk I/O operations and also massive resources for communication and storage.
Apache Spark \cite{ex8} is another good cloud platform that is suitable for data mining.
In comparison with Hadoop, a Resilient Distributed Datasets (RDD) model and a Directed Acyclic Graph (DAG) model built on a memory computing framework is supported for Spark.
It allows us to store a data cache in memory and to perform computation and iteration for the same data directly from memory.
The Spark platform saves huge amounts of disk I/O operation time.
Therefore, it is more suitable for data mining with iterative computation.

The Random Forest (RF) algorithm \cite{ex9} is a suitable data mining algorithm for big data.
It is an ensemble learning algorithm using feature sub-space to construct the model.
Moreover, all decision trees can be trained concurrently, hence it is also suitable for parallelization.

\subsection{Our Contributions}
In this paper, we propose a Parallel Random Forest (PRF) algorithm for big data that is implemented on the Apache Spark platform.
The PRF algorithm is optimized based on a hybrid approach combining data-parallel and task-parallel optimization.
To improve the classification accuracy of PRF, an optimization is proposed prior to the parallel process.
Extensive experiment results indicate the superiority of PRF and depict its significant advantages over other algorithms in terms of the classification accuracy and performance.
Our contributions in this paper are summarized as follows.

\begin{itemize}
  \item An optimization approach is proposed to improve the accuracy of PRF, which includes a dimension-reduction approach in the training process and a weighted voting method in the prediction process.
  \item A hybrid parallel approach of PRF is utilized to improve the performance of the algorithm, combining data-parallel and task-parallel optimization. In the data-parallel optimization, a vertical data-partitioning method and a data-multiplexing method are performed.
  \item Based on the data-parallel optimization, a task-parallel optimization is proposed and implemented on Spark. A training task DAG of PRF is constructed based on the RDD model, and different task schedulers are invoked to perform the tasks in the DAG. The performance of PRF is improved noticeably.
\end{itemize}

The rest of the paper is organized as follows.
Section 2 reviews the related work.
Section 3 gives the RF algorithm optimization from two aspects.
The parallel implementation of the RF algorithm on Spark is developed in Section 4.
Experimental results and evaluations are shown in Section 5 with respect to the classification accuracy and performance.
Finally, Section 6 presents a conclusion and future work.

\section{Related Work}
Although traditional data processing techniques have achieved good performance for small-scale and low-dimensional datasets, they are difficult to be applied to large-scale data efficiently \cite{ex10,ex11,ex12}.
When a dataset becomes more complex with characteristics of a complex structure, high dimensionality, and a large size, the accuracy and performance of traditional data mining algorithms are significantly declined \cite{ex13}.

Due to the need to address the high-dimensional and noisy data, various improvement methods have been introduced by researchers.
Xu \cite{ex14} proposed a dimension-reduction method for the registration of high-dimensional data.
The method combines datasets to obtain an image pair with a detailed texture and results in improved image registration.
Tao et al. \cite{ex15} and Lin et al. \cite{ex16} introduced some classification algorithms for high-dimensional data to address the issue of dimension-reduction.
These algorithms use multiple kernel learning framework and multilevel maximum margin features and achieve efficient dimensionality reduction in binary classification problems.
Strobl \cite{ex17} and Bernard \cite{ex18} studied the variable importance measures of RF and proposed some improved models for it.
Taghi et al. \cite{ex19} compared the boosting and bagging techniques and proposed an algorithm for noisy and imbalanced data.
Yu et al. \cite{ex20} and Biau \cite{ex21} focused on RF for high-dimensional and noisy data and applied RF in many applications such as multi-class action detection and facial feature detection, and achieved a good effort.
Based on the existing research results, we propose a new optimization approach in this paper to address the problem of high-dimensional and noisy data, which reduces the dimensionality of the data according to the structure of the RF and improves the algorithm's accuracy with a low computational cost.

Focusing on the performance of classification algorithms for large-scale data, numerous studies on the intersection of parallel/distributed computing and the learning of tree models were proposed.
Basilico et al. \cite{ex22} proposed a COMET algorithm based on MapReduce, in which multiple RF ensembles are built on distributed blocks of data.
Svore et al. \cite{ex23} proposed a boosted decision tree ranking algorithm, which addresses the speed and memory constraints by distributed computing.
Panda et al. \cite{ex24} introduced a scalable distributed framework based on MapReduce for the parallel learning of tree models over large datasets.
A parallel boosted regression tree algorithm was proposed in \cite{ex25} for web search ranking, in which a novel method for parallelizing the training of GBRT was performed based on data partitioning and distributed computing.

Focusing on resource allocation and task-parallel execution in a parallel and distributed environment,
Warneke et al. \cite{ex26} implemented a dynamic resource allocation for efficient parallel data processing in a cloud environment.
Lena et al. \cite{ex27} carried out an energy-aware scheduling of MapReduce jobs for big data applications.
Luis et al. \cite{ex28} proposed a robust resource allocation of data processing on a heterogeneous parallel system, in which the arrival time of datasets are uncertainty.
Zhang et al. \cite{ex29} proposed an evolutionary scheduling of dynamic multitasking workloads for big data analysis in an elastic cloud.
Meanwhile, our team also focused on parallel tasks scheduling on heterogeneous cluster and distributed systems and achieved positive results\cite{ex30,ex31}.

Apache Spark Mllib \cite{ex32} parallelized the RF algorithm (referred to Spark-MLRF in this paper) based on a data-parallel optimization to improve the performance of the algorithm.
However, there exist many drawbacks in the Spark-MLRF.
First, in the stage of determining the best split segment for continuous features, a method of sampling for each partition of the dataset is used to reduce the data transmission operations.
However, the cost of this method is its reduced accuracy.
In addition, because the data-partitioning method in Spark-MLRF is a horizontal partition, the data communication of the feature variable gain ratio computing is a global communication.

To improve the performance of the RF algorithm and mitigate the data communication cost and workload imbalance problems of large-scale data in parallel and distributed environments, we propose a hybrid parallel approach for RF combining data-parallel and task-parallel optimization based on the Spark RDD and DAG models.
In comparison with the existing study results, our method reduces the volume of the training dataset without decreasing the algorithm's accuracy.
Moreover, our method mitigates the data communication and workload imbalance problems of large-scale data in parallel and distributed environments.

\section{Random Forest Algorithm Optimization}
Owing to the improvement of the classification accuracy for high-dimensional and large-scale data, we propose an optimization approach for the RF algorithm.
First, a dimension-reduction method is performed in the training process.
Second, a weighted voting method is constructed in the prediction process.
After these optimizations, the classification accuracy of the algorithm is evidently improved.

\subsection{Random Forest Algorithm}
The random forest algorithm is an ensemble classifier algorithm based on the decision tree model.
It generates $k$ different training data subsets from an original dataset using a bootstrap sampling approach, and then, $k$ decision trees are built by training these subsets.
A random forest is finally constructed from these decision trees.
Each sample of the testing dataset is predicted by all decision trees, and the final classification result is returned depending on the votes of these trees.

\begin{figure}[!ht]
\setlength{\abovecaptionskip}{0pt}
\setlength{\belowcaptionskip}{0pt}
\centering
\includegraphics[width=2.5in]{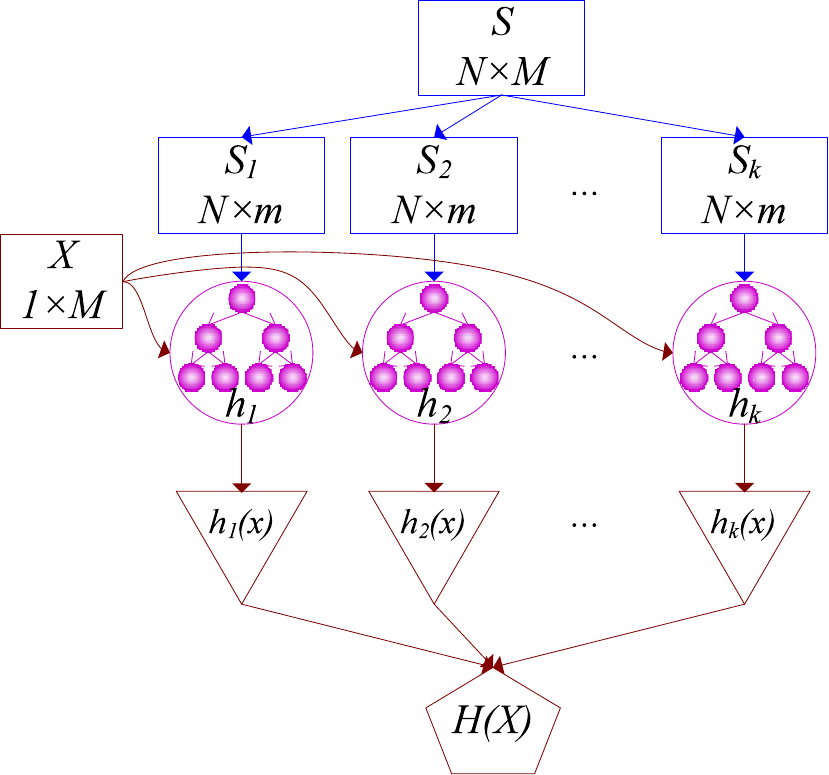}
\caption{Process of the construction of the RF algorithm}
\label{fig1}
\end{figure}

The original training dataset is formalized as $S=\{(x_{i},y_{j}),i=1,2,...,N; j=1,2,...,M\}$, where $x$ is a sample and $y$ is a feature variable of $S$.
Namely, the original training dataset contains $N$ samples, and there are $M$ feature variables in each sample.
The main process of the construction of the RF algorithm is presented in Fig. \ref{fig1}.

The steps of the construction of the random forest algorithm are as follows.

\textbf{Step 1.} Sampling $k$ training subsets.

In this step, $k$ training subsets are sampled from the original training dataset $S$ in a bootstrap sampling manner.
Namely, $N$ records are selected from $S$ by a random sampling and replacement method in each sampling time.
After the current step, $k$ training subsets are constructed as a collection of training subsets $S_{Train}$:

\begin{center}
$S_{Train} =\{S_{1}, S_{2},..., S_{k}\}$.
\end{center}

At the same time, the records that are not to be selected in each sampling period are composed as an Out-Of-Bag (OOB) dataset.
In this way, $k$ OOB sets are constructed as a collection of $S_{OOB}$:

\begin{center}
$S_{OOB} =\{OOB_{1}, OOB_{2},..., OOB_{k}\}$,
\end{center}
where $k \ll N$, $S_{i} \bigcap OOB_{i} = \phi$ and $S_{i} \bigcup OOB_{i} = S$.
To obtain the classification accuracy of each tree model, these OOB sets are used as testing sets after the training process.

\textbf{Step 2.} Constructing each decision tree model.

In an RF model, each meta decision tree is created by a C4.5 or CART algorithm from each training subset $S_{i}$.
In the growth process of each tree, $m$ feature variables of dataset $S_{i}$ are randomly selected from $M$ variables.
In each tree node's splitting process, the gain ratio of each feature variable is calculated, and the best one is chosen as the splitting node.
This splitting process is repeated until a leaf node is generated.
Finally, $k$ decision trees are trained from $k$ training subsets in the same way.

\textbf{Step 3.} Collecting $k$ trees into an RF model.

The $k$ trained trees are collected into an RF model, which is defined in Eq. (\ref{equ2}):

\begin{equation}
\label{equ2}
H(X, \Theta j) =\sum_{i=1}^{k} {h_{i}(x, \Theta j)}, (j=1,2,...,m),
\end{equation}
where $h_{i}(x, \Theta j)$ is a meta decision tree classifier, $X$ are the input feature vectors of the training dataset, and $\Theta j$ is an independent and identically distributed random vector that determines the growth process of the tree.

\subsection{Dimension Reduction for High-Dimensional Data}
\begin{figure}[!ht]
\setlength{\abovecaptionskip}{0pt}
\setlength{\belowcaptionskip}{0pt}
\centering
\includegraphics[width=3.0in]{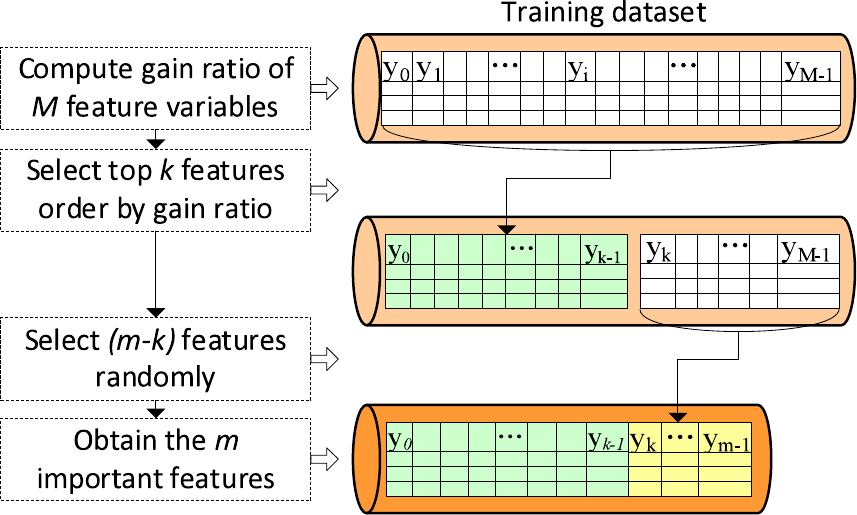}
\caption{Dimension-reduction in the training process}
\label{fig2}
\end{figure}

To improve the accuracy of the RF algorithm for the high-dimensional data, we present a new dimension-reduction method to reduce the number of dimensions according to the importance of the feature variables.
In the training process of each decision tree, the Gain Ratio (GR) of each feature variable of the training subset is calculated and sorted in descending order.
The top $k$ variables ($k\ll M$) in the ordered list are selected as the principal variables, and then, we randomly select $(m-k)$ further variables from the remaining $(M-k)$ ones.
Therefore, the number of dimensions of the dataset is reduced from $M$ to $m$.
The process of dimension-reduction is presented in Fig. \ref{fig2}.

First, in the training process of each decision tree, the entropy of each feature variable is calculated prior to the node-splitting process.
The entropy of the target variable in the training subset $S_{i}$ ($i=1,2,...,k$) is defined in Eq. (\ref{equ5}):

\begin{equation}
\label{equ5}
Entropy(S_{i} )=\sum_{a=1}^{c1}-p_{a}\log p_{a},
\end{equation}
where $c1$ is the number of different values of the target variable in $S_{i}$, and $p_{a}$ is the probability of the type of value $a$ within all types in the target variable subset.

Second, the entropy for each input variable of $S_{i}$, except for the target variable, is calculated.
The entropy of each input variable $y_{ij}$ is defined in Eq. (\ref{equ6}):

\begin{equation}
\label{equ6}
Entropy(y_{ij} )=\sum_{v\in V(y_{ij})} \frac{|S_{(v,i)}|}{|S_{i}|} Entropy(v(y_{ij})),
\end{equation}
where the elements of Eq. (\ref{equ6}) are described in Table \ref{table31}.

\begin{table}[!ht]
\setlength{\abovecaptionskip}{0pt}
\setlength{\belowcaptionskip}{0pt}
\renewcommand{\arraystretch}{1.3}
\caption{Explanation of the elements of Eq. (\ref{equ6})}
\centering
\label{table31}
\tabcolsep1pt
\begin{tabular*}{3.5in}{c l}
\hline
Element & Description\\
\hline
 $y_{ij}$ & the $j$-th feature variable of $S_{i}$, $j=1,2,..., M$.\\

 $V(y_{ij})$ & the set of all possible values of $y_{ij}$.\\

 $|S_{i}|$ & the number of samples in $S_{i}$.\\

 $S_{(v,i)}$ & a sample subset in $S_{i}$, where the value of $y_{j}$ is $v$.\\

 $|S_{(v,i)}|$ & the number of the sample subset $S_{(v,i)}$.\\
\hline
\end{tabular*}
\end{table}

Third, the self-split information $I(y_{ij})$ of each input variable is calculated, as defined in Eq. (\ref{equ7}):

\begin{equation}
\label{equ7}
I(y_{ij})=\sum_{a=1}^{c2} -p_{(a,j)} \log_{2}(p_{(a,j)}),
\end{equation}
where $c2$ is the number of different values of $y_{ij}$, and $p_{(a,j)}$ is the probability of the type of value $a$ within all types in variable $y_{j}$.
Then, the information gain of each feature variable is calculated, as defined in Eq. (\ref{equ8}):

\begin{equation}
\label{equ8}
\begin{aligned}
G(y_{ij})=&Entropy(S_{i})-Entropy(y_{ij})\\
=&Entropy(S_{i})\\
&-\sum_{v\in V(y_{ij})}\frac{|S_{(v,i)}|}{|S_{i}|} Entropy(v(y_{ij})),
\end{aligned}
\end{equation}
where $v(y_{j})\in V(y_{j})$.

By using the information gain to measure the feature variables, the largest value is selected easily, but it will lead to an over fitting problem.
To overcome this problem, a gain ratio value is taken to measure the feature variables, and the features with the maximum value are selected.
The gain ratio of the feature variable $y_{ij}$ is defined in Eq. (\ref{equ9}):

\begin{equation}
\label{equ9}
GR(y_{ij}) = \frac{G(y_{ij})}{I(y_{ij})}.
\end{equation}

To reduce the dimensions of the training dataset, we calculate the importance of each feature variable according to the gain ratio of the variable.
Then, we select the most important features and delete the ones with less importance.
The importance of each feature variable is defined as follows.

\textbf{Definition 1.} The importance of each feature variable in a training subset refers to the portion of the gain ratio of the feature variable compared with the total feature variables.
The importance of feature variable $y_{ij}$ is defined as $VI(y_{ij})$ in Eq. (\ref{equ10}):

\begin{equation}
\label{equ10}
VI(y_{ij})=\frac{GR(y_{ij})}{\sum_{(a=1)}^{M}GR(y_{(i,a)})}.
\end{equation}

The importance values of all feature variables are sorted in descending order, and the top $k$ ($k\ll M, k < m $) values are selected as the most important.
We then randomly select $(m-k)$ further feature variables from the remaining $(M-k)$ ones.
Thus, the number of dimensions of the dataset is reduced from $M$ to $m$.
Taking the training subset $S_{i}$ as an example, the detailed steps of the dimension-reduction in the training process are presented in Algorithm \ref{alg1}.

\begin{algorithm}[!ht]
\caption{Dimension-reduction in the training process}
\label{alg1}
\begin{algorithmic}[1]
{\small
\REQUIRE ~\\
    $S_{i}$: the $i$th training subset;\\
    $k$: the number of important variables selected by $VI$;\\
    $m$: the number of the selected feature variables.\\
\ENSURE ~\\
    $F_{i}$: a set of $m$ important feature variables of $S_{i}$.
\STATE create an empty string array $F_{i}$;
\STATE calculate $Entropy(S_{i})$ for the target feature variable;
\FOR{each feature variable $y_{ij}$ in $S_{i}$}
\STATE calculate $Entropy(y_{ij})$ for each input feature variable;
\STATE calculate gain $G(y_{ij}) \leftarrow Entropy(S_{i}) - Entropy(y_{ij})$;
\STATE calculate split information $I(y_{ij}) \leftarrow \sum_{a=1}^{c2} -p_{(a,j)} \log_{2}(p_{(a,j)})$;
\STATE calculate gain ratio $GR(y_{ij}) \leftarrow \frac{G(y_{ij})}{I(y_{ij})}$;
\ENDFOR
\STATE calculate variable importance $VI(y_{ij}) \leftarrow \frac{GR(y_{ij})}{\sum_{(a=1)}^{M}GR(y_{(i,a)})}$ for feature variable $y_{ij}$;
\STATE sort $M$ feature variables in descending order by $VI(y_{ij})$;
\STATE put top $k$ feature variables to $F_{i}[0,...,k-1]$;
\STATE set $c \leftarrow 0$;
\FOR {$j=k$ to $M-1$}
      \WHILE{$c < (m-k)$}
      \STATE select $y_{ij}$ from $(M-k)$ randomly;
      \STATE put $y_{ij}$ to $F_{i}[k+c]$;
      \STATE $c$ $\leftarrow$ $c+1$;
      \ENDWHILE
\ENDFOR
\RETURN $F_{i}$.
}
\end{algorithmic}
\end{algorithm}

In comparison with the original RF algorithm, our dimension-reduction method ensures that the $m$ selected feature variables are optimal while maintaining the same computational complexity as the original algorithm.
This balances the accuracy and diversity of the feature selection ensemble of the RF algorithm and prevents the problem of classification over fitting.

\subsection{Weighted Voting Method}
In the prediction and voting process, the original RF algorithm uses a traditional direct voting method.
In this case, if the RF model contains noisy decision trees, it likely leads to a classification or regression error for the testing dataset.
Consequently, its accuracy is decreased.
To address this problem, a weighted voting approach is proposed in this section to improve the classification accuracy for the testing dataset.
The accuracy of the classification or regression of each tree is regarded as the voting weight of the tree.

After the training process, each OOB set $OOB_{i}$ is tested by its corresponding trained tree $h_{i}$.
Then, the classification accuracy $CA_{i}$ of each decision tree $h_{i}$ is computed.

\textbf{Definition 2.} The classification accuracy of a decision tree is defined as the ratio of the average number of votes in the correct classes to that in all classes, including error classes, as classified by the trained decision tree.
The classification accuracy is defined in Eq. (\ref{equ11}):

\begin{equation}
\label{equ11}
CA_{i}=\frac{I(h_{i}(x)=y)}{I(h_{i}(x)=y)+ \sum{I(h_{i}(x)=z)}},
\end{equation}
where $I(\cdot)$ is an indicator function, $y$ is a value in the correct class, and $z$ is a value in the error class ($z\neq y$).

In the prediction process, each record of the testing dataset $X$ is predicted by all decision trees in the RF model, and then, a final vote result is obtained for the testing record.
When the target variable of $X$ is quantitative data, the RF is trained as a regression model.
The result of the prediction is the average value of $k$ trees.
The weighted regression result $H_{r}(X)$ of $X$ is defined in Eq. (\ref{equ3}):

\begin{equation}
\label{equ3}
\begin{aligned}
H_{r}(X) &=\frac{1}{k} \sum_{i=1}^{k} {[w_{i}\times h_{i}(x)]}\\
&=\frac{1}{k}\sum_{i=1}^{k}{[CA_{i}\times h_{i}(x)]},
\end{aligned}
\end{equation}
where $w_{i}$ is the voting weight of the decision tree $h_{i}$.

Similarly, when the target feature of $X$ is qualitative data, the RF is trained as a classification model.
The result of the prediction is the majority vote of the classification results of $k$ trees.
The weighted classification result $H_{c}(X)$ of $X$ is defined in Eq. (\ref{equ4}):

\begin{equation}
\label{equ4}
\begin{aligned}
H_{c}(X)&=Majority_{i=1}^{k}{[ w_{i}\times h_{i}(x)]}\\
&=Majority_{i=1}^{k}{[CA_{i}\times h_{i}(x)]}.
\end{aligned}
\end{equation}

The steps of the weighted voting method in the prediction process are described in Algorithm \ref{alg3}.

\begin{algorithm}[!ht]
\caption{Weighted voting in the prediction process}
\label{alg3}
\begin{algorithmic}[1]
{\small
\REQUIRE ~\\
    $X$: a testing dataset;\\
    $PRF_{trained}$: the trained PRF model.\\
\ENSURE ~\\
    $H(X)$: the final prediction result for $X$.\\
\FOR {each testing data $x$ in $X$}
\FOR{each decision tree $T_{i}$ in $PRF_{trained}$}
\STATE predict the classification or regression result $h_{i}(x)$ by $T_{i}$;
\ENDFOR
\ENDFOR
 \STATE set classification accuracy $CA_{i}$ as the weight $w_{i}$ of $T_{i}$;
\FOR {each testing data $x$ in $X$}
 \IF {(operation type == classification)}
      \STATE vote the final result $H_{c}(x) \leftarrow Majority_{i=1}^{k}{[ w_{i}\times h_{i}(x)]}$;
      \STATE $H(X) \leftarrow H_{c}(x)$;
      \ELSIF {(operation type == regression)}
      \STATE vote the final result $H_{r}(x) \leftarrow \frac{1}{k} \sum_{i=1}^{k} {[w_{i}\times h_{i}(x)]}$;
      \STATE $H(X) \leftarrow H_{r}(x)$;
      \ENDIF
\ENDFOR
\RETURN $H(X)$.
}
\end{algorithmic}
\end{algorithm}

In the weighted voting method of RF, each tree classifier corresponds to a specified reasonable weight for voting on the testing data.
Hence, this improves the overall classification accuracy of RF and reduces the generalization error.

\subsection{Computational Complexity}
The computational complexity of the original RF algorithm is $O(kMN \log N)$, where $k$ is the number of decision trees in RF, $M$ is the number of features, $N$ is the number of samples, and $\log N$ is the average depth of all tree models.
In our improved PRF algorithm with dimension-reduction (PRF-DR) described in Section 3, the time complexity of the dimension reduction is $O(MN)$.
The computational complexity of the splitting process for each tree node is set as one unit (1), which contains functions such as $entropy()$, $gain()$, and $gainratio()$ for each feature subspace.
After the dimension reduction, the number of features is reduced from $M$ to $m$ ($m \ll M$).
Therefore, the computational complexity of training a meta tree classifier is $O(MN+mN \log N)$, and the total computational complexity of the PRF-DR algorithm is $O(k(MN+mN \log N))$.

\section{Parallelization of the Random Forest Algorithm on Spark}
To improve the performance of the RF algorithm and mitigate the data communication cost and workload imbalance problems of large-scale data in a parallel and distributed environment, we propose a Parallel Random Forest (PRF) algorithm on Spark.
The PRF algorithm is optimized based on a hybrid parallel approach combining data-parallel and task-parallel optimization.
From the perspective of data-parallel optimization, a vertical data-partitioning method and a data-multiplexing method are performed.
These methods reduce the volume of data and the number of data transmission operations in the distributed environment without reducing the accuracy of the algorithm.
From the perspective of task-parallel optimization, a dual-parallel approach is carried out in the training process of the PRF algorithm, and a task DAG is created according to the dependence of the RDD objects.
Then, different task schedulers are invoked to perform the tasks in the DAG.
The dual-parallel training approach maximizes the parallelization of PRF and improves the performance of PRF.
Then task schedulers further minimize the data communication cost among the Spark cluster and achieve a better workload balance.

\subsection{Data-Parallel Optimization}
We introduce the data-parallel optimization of the PRF algorithm, which includes a vertical data-partitioning and a data-multiplexing approach.
First, taking advantage of the RF algorithm's natural independence of feature variables and the resource requirements of computing tasks, a vertical data-partitioning method is proposed for the training dataset.
The training dataset is split into several feature subsets, and each feature subset is allocated to the Spark cluster in a static data allocation way.
Second, to address the problem that the data volume increases linearly with the increase in the scale of RF, we present a data-multiplexing method for PRF by modifying the traditional sampling method.
Notably, our data-parallel optimization method reduces the volume of data and the number of data transmission operations without reducing the accuracy of the algorithm.
The increase in the scale of the PRF does not lead to a change in the data size and storage location.

\subsubsection{Vertical Data Partitioning}
In the training process of PRF, the gain-ratio computing tasks of each feature variable take up most of the training time.
However, these tasks only require the data of the current feature variable and the target feature variable.
Therefore, to reduce the volume of data and the data communication cost in a distributed environment, we propose a vertical data-partitioning approach for PRF according to the independence of feature variables and the resource requirements of computing tasks.
The training dataset is divided into several feature subsets.

\begin{figure}[!ht]
\setlength{\abovecaptionskip}{0pt}
\setlength{\belowcaptionskip}{0pt}
\centering
\includegraphics[width=1.8in]{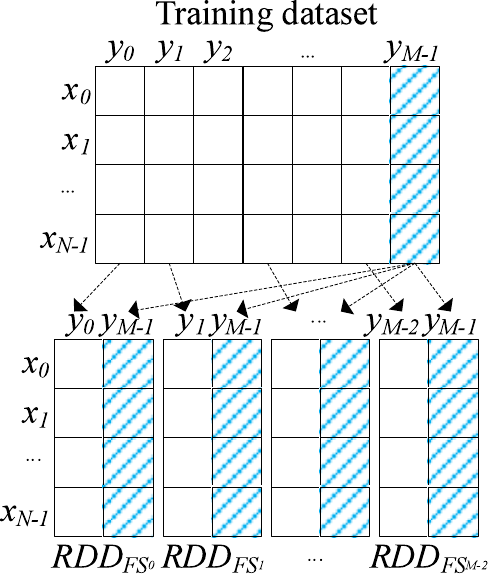}
\caption{Process of the vertical data-partitioning method}
\label{fig3}
\end{figure}

Assume that the size of training dataset $S$ is $N$ and there are $M$ feature variables in each record.
$y_{0} \sim y_{M-2}$ are the input feature variables, and $y_{M-1}$ is the target feature variable.
Each input feature variable $y_{j}$ and the variable $y_{M-1}$ of all records are selected and generated to a feature subset $FS_{j}$, which is represented as:

\begin{center}
$FS_{j}=
\left[
\begin{array}{c}
<0, y_{0j}, y_{0(M-1)}>,\\
<1, y_{1j}, y_{1(M-1)}>,\\
...,\\
<i, y_{ij}, y_{i(M-1)}>,\\
...,\\
<(N-1), y_{(N-1)j}, y_{(N-1)(M-1)}>
 \end{array}
\right]$,
\end{center}
where $i$ is the index of each record of the training dataset $S$, and $j$ is the index of the current feature variable.
In such a way, $S$ is split into $(M-1)$ feature subsets before dimension-reduction.
Each subset is loaded as an RDD object and is independent of the other subsets.
The process of the vertical data-partitioning is presented in Fig. \ref{fig3}.

\subsubsection{Data-Multiplexing Method}
To address the problem that the volume of the sampled training dataset increases linearly with the increase in the RF scale, we present a data-multiplexing method for PRF by modifying the traditional sampling method.
In each sampling period, we do not copy the sampled data but just note down their indexes into a Data-Sampling-Index (DSI) table.
Then, the DSI table is allocated to all slave nodes together with the feature subsets.
The computing tasks in the training process of each decision tree load the corresponding data from the same feature subset via the DSI table.
Thus, each feature subset is reused effectively, and the volume of the training dataset will not increase any more despite the expansion of the RF scale.

First, we create a DSI table to save the data indexes generated in all sampling times.
As mentioned in Section 3.1, the scale of a RF model is $k$.
Namely, there are $k$ sampling times for the training dataset, and $N$ data indexes are noted down in each sampling time.
An example of the DSI table of PRF is presented in Table \ref{table41}.

\begin{table}[!ht]
\renewcommand{\arraystretch}{1.3}
\setlength{\abovecaptionskip}{0pt}
\setlength{\belowcaptionskip}{0pt}
\caption{Example of the DSI table of PRF}
\centering
\label{table41}
\begin{tabular*}{3.0in}{c c c c c c c c c}
\hline
& & \multicolumn{7}{c}{Data indexes of training dataset} \\
\hline
\multirow{6}{*}{\begin{sideways}{Sampling times}\end{sideways}} &
    $Sample_{0}$   &   1  &   3   &   8   &   5   &   10  &   0   &   ...  \\
&   $Sample_{1}$   &   2  &   4   &   1   &   9   &   7   &   8   &   ...  \\
&   $Sample_{2}$   &   9  &   1   &   12  &   92  &   2   &   5   &   ...  \\
&   $Sample_{3}$   &   3  &   8   &   87  &   62  &   0   &   2   &   ...   \\
&   ...         & ...  &  ...  &  ...  &   ... &  ...  &  ...  &   ...   \\
&   $Sample_{k-1}$   &   9  &   1   &   4   &   43  &   3   &   5   &   ...    \\
\hline
\end{tabular*}
\end{table}

Second, the DSI table is allocated to all slave nodes of the Spark cluster together with all feature subsets.
In the subsequent training process, the gain-ratio computing tasks of different trees for the same feature variable are dispatched to the slaves where the required subset is located.

\begin{figure}[!ht]
\setlength{\abovecaptionskip}{0pt}
\setlength{\belowcaptionskip}{0pt}
\centering
\includegraphics[width=2.5in]{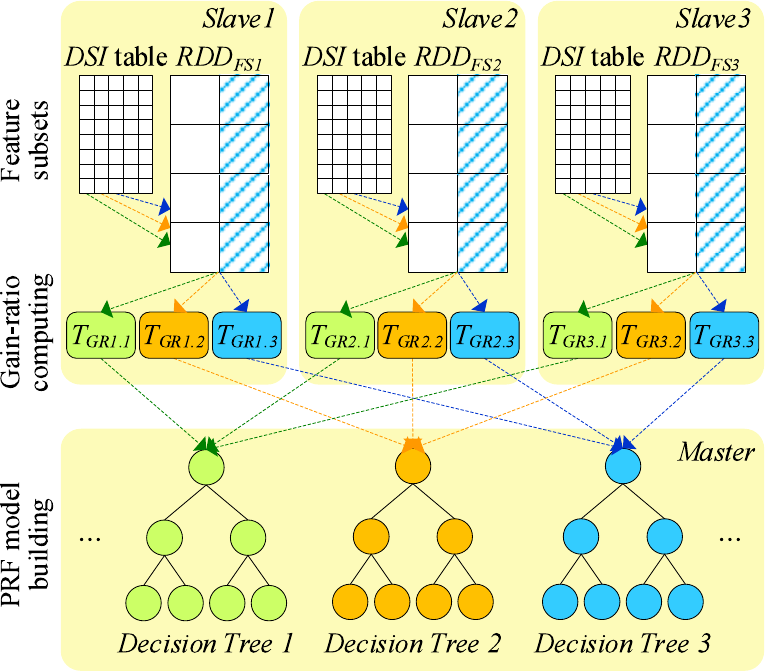}
\caption{Process of the data-multiplexing method of PRF}
\label{fig4}
\end{figure}

Third, each gain-ratio computing task accesses the relevant data indexes from the DSI table, and obtains the feature records from the same feature subset according to these indexes.
The process of the data-multiplexing method of PRF is presented in Fig. \ref{fig4}.

In Fig. \ref{fig4}, each $RDD_{FS}$ refers to an RDD object of a feature subset, and each $T_{GR}$ refers to a gain-ratio computing task.
For example, we allocate tasks $\{T_{GR1.1}, T_{GR1.2}, T_{GR1.3}\}$ to $Slave1$ for the feature subset $RDD_{FS1}$, allocate tasks $\{T_{GR2.1}, T_{GR2.2}, T_{GR2.3}\}$ to $Slave2$ for $RDD_{FS2}$, and allocate tasks $\{T_{GR3.1}, T_{GR3.2}, T_{GR3.3}\}$ to $Slave3$ for $RDD_{FS3}$.
From the perspective of the decision trees, tasks in the same slave node belong to different trees.
For example, tasks $T_{GR1.1}$, $T_{GR1.2}$, and $T_{GR1.3}$ in the $Slave1$ belong to ``$DecisionTree1$", ``$DecisionTree2$", and ``$DecisionTree3$", respectively.
These tasks obtain records from the same feature subset according to the corresponding indexes in DSI, and compute the gain ratio of the feature variable for different decision trees.
After that, the intermediate results of these tasks are submitted to the corresponding subsequent tasks to build meta decision trees.
Results of the tasks $\{T_{GR1.1}$, $T_{GR2.1}$, $T_{GR3.1}\}$ are combined for the tree node splitting process of ``$DecisionTree1$", and results of the tasks $\{T_{GR1.2}$, $T_{GR2.2}$, $T_{GR3.2}\}$ are combined for that of ``$DecisionTree2$".

\subsubsection{Static Data Allocation}
\begin{figure}[!ht]
\setlength{\abovecaptionskip}{0pt}
\setlength{\belowcaptionskip}{0pt}
\centering
\includegraphics[width=3.0in]{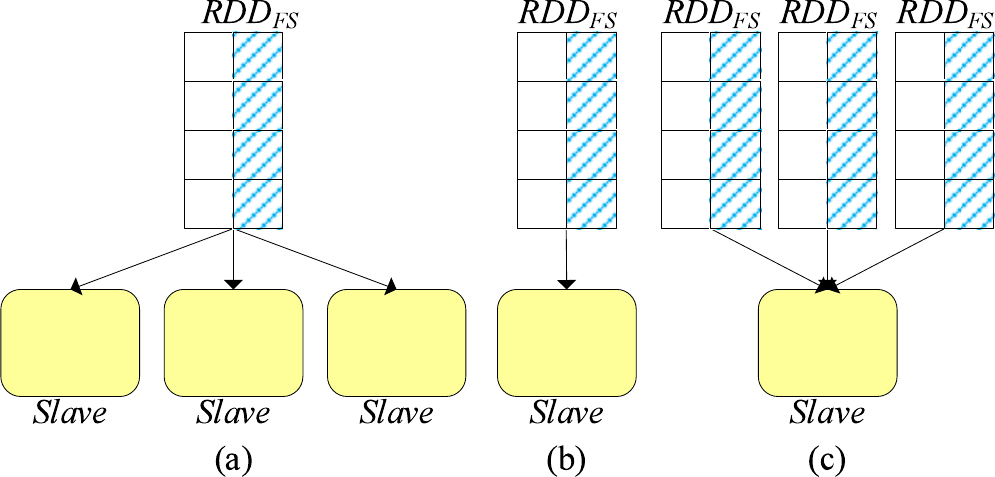}
\caption{Example of 3 scenarios of the data allocation}
\label{fig5}
\end{figure}

To achieve a better balance of data storage and workload, after the vertical data-partitioning, a static data allocation method is applied for the feature subsets.
Namely, these subsets are allocated to a distributed Spark cluster before the training process of PRF.
Moreover, because of the difference of the data type and volume of each feature subset, the workloads of the relevant subsequent computing tasks will be different as well.
As we know, a Spark cluster is constructed by a master node and several slave nodes.
We define our allocation function to determine each feature subset be allocated to which nodes, and allocate each feature subset according to its volume.
There are 3 scenarios in the data allocation scheme.
Examples of the 3 scenarios of the data allocation method are shown in Fig. \ref{fig5}.

In Fig. \ref{fig5}, (a) when the size of a feature subset is greater than the available storage capacity of a slave node, this subset is allocated to limited multiple slaves that have similar physical locations.
(b) When the size of a feature subset is equal to the available storage capacity of a slave node, the subset is allocated to the node.
(c) When the size of a feature subset is smaller than the available storage capacity of a slave node, this node will accommodate multiple feature subsets.
In case (a), the data communication operations of the subsequent gain-ratio computing tasks occur among the slave nodes where the current feature subset is located.
These data operations are local communications but not global communications.
In cases (b) and (c), no data communication operations occur among different slave nodes in the subsequent gain-ratio computation process.
The steps of the vertical data-partitioning and static data allocation of PRF are presented in Algorithm \ref{alg41}.

\begin{algorithm}
\caption{Vertical data-partitioning and static data allocation of PRF}
\label{alg41}
\begin{algorithmic}[1]
\vspace{.2cm}
{\small
\REQUIRE ~\\
     $RDD_{o}$: an RDD object of the original training dataset $S$.
\ENSURE ~\\
    $L_{FS}$: a list of the indexes of each feature subset's RDD object and the allocated slave nodes.
\FOR {$j=0$ to $(M-2)$}
  \STATE $RDD_{FSj} \leftarrow$ $RDD_{o}$.\textbf{map}
  \STATE \quad $<i, y_{ij}, y_{i(M-1)}> \leftarrow RDD_{o}$.verticalPartition(j);
  \STATE \textbf{end map}.collect()
  \STATE $slaves \leftarrow$ findAvailableSlaves().sortbyIP();
  \IF {$RDD_{FSj}$.size $<$ $slaves[0]$.availablesize}
  \STATE dataAllocation($RDD_{FSj}$, $slaves[0]$);
  \STATE $slaves[0]$.availablesize $\leftarrow$ $slaves[0]$.availablesize - $RDD_{FSj}$.size;
  \STATE $L_{FS}$ $\leftarrow$ $<RDD_{FSj}.id, slaves[0].nodeid>$;
  \ELSE
  \WHILE {$RDD_{FSj} \neq null$}
  \STATE ($RDD_{FSj}^{'}$, $RDD_{FSj}^{''})$ $\leftarrow$ dataPartition($RDD_{FSj}$, $slaves[i]$.availablesize));
  \STATE dataAllocation($RDD_{FSj}^{'}$, $slaves[i]$);
  \STATE $RDD_{FSj}^{'}$.persist();
  \STATE $slaves[i]$.availablesize $\leftarrow$ $slaves[i]$.availablesize - $RDD_{FSj}^{'}$.size;
  \STATE $slavesids$ $\leftarrow$ $slaves[i].nodeid$;
  \STATE $RDD_{FSj} \leftarrow RDD_{FSj}^{''}$;
  \STATE $i \leftarrow i+1$;
  \ENDWHILE
  \STATE $L_{FS}$ $\leftarrow$ $<RDD_{FSj}.id, slavesids>$;
  \ENDIF
  \ENDFOR
  \RETURN $L_{FS}$.
  }
\end{algorithmic}
\end{algorithm}

In Algorithm \ref{alg41}, $RDD_{o}$ is split into $(M-1)$ $RDD_{FS}$ objects via the vertical data-partitioning function firstly.
Then, each $RDD_{FS}$ is allocated to slave nodes according to its volume and the available storage capacity of the slave nodes.
To reuse the training dataset, each RDD object of the feature subset is allocated and persisted to Spark cluster via a $dataAllocation()$ function and a $persist()$ function.

\subsection{Task-Parallel Optimization}
Each decision tree of PRF is built independent of each other, and each sub-node of a decision tree is also split independently.
The structures of the PRF model and decision tree model make the computing tasks have natural parallelism.
Based on the results of the data-parallel optimization, we propose a task-parallel optimization for PRF and implement it on Spark.
A dual-parallel approach is carried out in the training process of PRF, and a task DAG is created according to the dual-parallel training process and the dependence of the RDD objects.
Then, different task schedulers are invoked to perform the tasks in the DAG.

\subsubsection{Parallel Training Process of PRF}
In our task-parallel optimization approach, a dual-parallel training approach is proposed in the training process of PRF on Spark.
$k$ decision trees of the PRF model are built in parallel at the first level of the training process.
And $(M-1)$ feature variables in each decision tree are calculated concurrently for tree node splitting at the second level of the training process.

There are several computing tasks in the training process of each decision tree of PRF.
According to the required data resources and the data communication cost, the computing tasks are divided into two types, gain-ratio computing tasks and node-splitting tasks, which are defined as follows.

\textbf{Definition 3.} Gain-ratio-computing task ($T_{GR}$) is a task set that is employed to compute the gain ratio of a feature variable from the corresponding feature subset, which includes a series of calculations for each feature variable, i.e., the entropy, the self-split information, the information gain, and the gain ratio.
The results of $T_{GR}$ tasks are submitted to the corresponding subsequent node-splitting tasks.

\textbf{Definition 4.} Node-splitting task ($T_{NS}$) is a task set that is employed to collect the results of the relevant $T_{GR}$ tasks and split the decision tree nodes, which includes a series of calculations for each tree node, such as determining the best splitting variable holds the highest gain ratio value and splitting the tree node by the variable.
After the tree node splitting, the results of $T_{NS}$ tasks are distributed to each slave to begin the next stage of the PRF's training process.

The steps of the parallel training process of the PRF model are presented in Algorithm \ref{alg6}.

\begin{algorithm}
\caption{Parallel training process of the PRF model}
\label{alg6}
\begin{algorithmic}[1]
\vspace{.2cm}
{\small
\REQUIRE ~\\
     $k$: the number of decision trees of the PRF model;\\
     $T_{DSI}$: the DSI table of PRF;\\
     $L_{FS}$: a list of the indexes of each feature subset's RDD object and the allocated slave nodes.\\
\ENSURE ~\\
     $PRF_{trained}$: the trained PRF model.\\
\FOR {$i$ = $0$ to $(k-1)$ }
\FOR {$j$ = $0$ to $(M-2)$}
\STATE  load feature subset $RDD_{FSj}$ $\leftarrow$ loadData($L_{FS}[i]$);
\\
 //$T_{GR}$:\\
\STATE $RDD_{(GR,best)} \leftarrow$ sc.parallelize($RDD_{FSj}$).\textbf{map}
\STATE $\;\;$ load sampled data $RDD_{(i,j)}$  $\leftarrow$ ($T_{DSI}[i]$, $RDD_{FSj}$);
\STATE $\;\;$ calculate the gain ratio $GR_{(i,j)} \leftarrow GR(RDD_{(i,j)})$;
\STATE \textbf{end map}
\\
 //$T_{NS}$:\\
\STATE $RDD_{(GR,best)} $.collect().sorByKey(GR).top(1);
\FOR {each value $y_{(j,v)}$ in $RDD_{(GR,best)}$ }
  \STATE split tree node $Node_{j} \leftarrow <y_{(j,v)},Value>$;
  \STATE append $Node_{j}$ to $T_{i}$;
\ENDFOR
\ENDFOR
\STATE $PRF_{trained}$ $\leftarrow$ $T_{i}$;
\ENDFOR
\RETURN $PRF_{trained}$.
}
\end{algorithmic}
\end{algorithm}

\begin{figure}[!ht]
\setlength{\abovecaptionskip}{0pt}
\setlength{\belowcaptionskip}{0pt}
\centering
\includegraphics[width=3.5in]{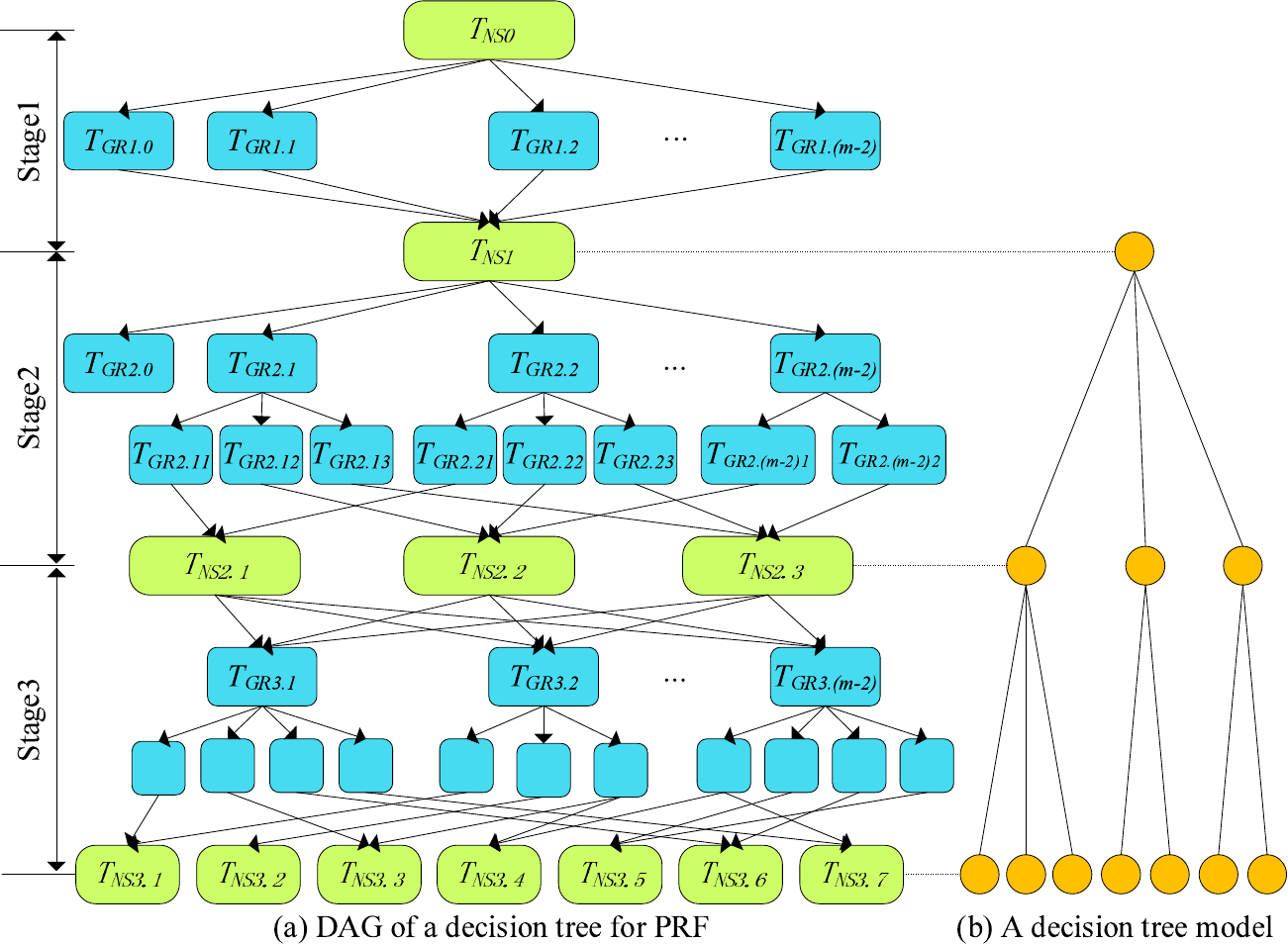}
\caption{Example of the task DAG for a decision tree of PRF}
\label{fig6}
\end{figure}

According to the parallel training process of PRF and the dependence of each RDD object, each job of the program of PRF's training process is split into different stages, and a task DAG is constructed with the dependence of these job stages.
Taking a decision tree model of PRF as an example, a task DAG of the training process is presented in Fig. \ref{fig6}.

There are several stages in the task DAG, which correspond to the levels of the decision tree model.
In stage 1, after the dimension-reduction, $(m-1)$ $T_{GR}$ tasks ($T_{GR1.0}$ $\sim$ $T_{GR1.(m-2)}$) are generated for the $(m-1)$ input feature variables.
These $T_{GR}$s compute the gain ratio the corresponding feature variable, and submit their results to $T_{NS1}$.
$T_{NS1}$ finds the best splitting variable and splits the first tree node for the current decision tree model.
Assuming that $y_{0}$ is the best splitting variable at the current stage, and the value of $y_{0}$ is in the range of $\{v_{01}, v_{02}, v_{03}\}$.
Hence, the first tree node is constructed by $y_{0}$, and 3 sub-nodes are split from the node, as shown in Fig. \ref{fig6}(b).
After tree node splitting, the intermediate result of $T_{NS1}$ are distributed to all slave nodes.
The result includes information of the splitting variable and the data index list of $\{v_{01}, v_{02}, v_{03}\}$.

In stage 2, because $y_{0}$ is the splitting feature, there is no $T_{GR}$ task for $FS_{0}$.
The potential workload balance problem of this issue will be discussed in Section \ref{WorkloadBalanceAnalysis}.
New $T_{GR}$ tasks are generated for all other feature subsets according to the result of $T_{NS1}$.
Due to the data index list of $\{v_{01}, v_{02}, v_{03}\}$, there are no more than 3 $T_{GR}$ tasks for each feature subset.
For example, tasks $T_{GR2.11}$, $T_{GR2.12}$, and $T_{GR2.13}$ calculate the data of $FS_{1}$ with the indexes corresponding to $v_{01}$, $v_{02}$, and $v_03$, respectively.
And the condition is similar in tasks for $FS_{2}$ $\sim$ $FS_{(m-2)}$.
Then, the results of tasks $\{T_{GR2.11}$, $T_{GR2.21}$, $T_{GR2.(m-2)1}\}$ are submitted to task $T_{NS2.1}$ for the same sub-tree-node splitting.
Tasks of other tree nodes and other stages are performed similarly.
In such a way, a task DAG of the training process of each decision tree model is built.
In addition, $k$ DAGs are built respectively for the $k$ decision trees of the PRF model.

\subsubsection{Task-Parallel Scheduling}
After the construction of the task DAGs of all the decision trees, the tasks in these DAGs are submitted to the Spark task scheduler.
There exist two types of computing tasks in the DAG, which have different resource requirements and parallelizables.
To improve the performance of PRF efficiently and further minimize the data communication cost of tasks in the distributed environment, we invoke two different task-parallel schedulers to perform these tasks.

In Spark, the $TaskSchedulerListener$ module monitors the submitted jobs, splits the job into different stages and tasks, and submits these tasks to the $TaskScheduler$ module.
The $TaskScheduler$ module receives the tasks and allocates and executes them using the appropriate executors.
According to the different allocations, the $TaskScheduler$ module includes 3 sub-modules, such as $LocalScheduler$, $ClusterScheduler$, and $MessosScheduler$.
Meanwhile, each task holds 5 types of locality property value: $PROCESS\_LOCAL$, $NODE\_LOCAL$, $NO\_PREF$, $PACK\_LOCAL$, and $ANY$.
We set the value of the locality properties of these two types of tasks and submit them into different task schedulers.
We invoke $LocalScheduler$ for $T_{GR}$ tasks and $ClusterScheduler$ to perform $T_{NS}$ tasks.

\textbf{(1) $LocalScheduler$ for $T_{GR}$ tasks.}

The $LocalScheduler$ module is a thread pool of the local computer, all tasks submitted by $DAGScheduler$ is executed in the thread pool, and the results will then be returned to $DAGScheduler$.
We set the locality property value of each $T_{GR}$ as $NODE\_LOCAL$ and submit it to a $LocalScheduler$ module.
In $LocalScheduler$, all $T_{GR}$ tasks of PRF are allocated to the slave nodes where the corresponding feature subsets are located.
These tasks are independent of each other, and there is no synchronization restraint among them.
If a feature subset is allocated to multiple slave nodes, the corresponding $T_{GR}$ tasks of each decision tree are allocated to these nodes.
And there exist local data communication operations of the tasks among these nodes.
If one or more feature subsets are allocated to one slave node, the corresponding $T_{GR}$ tasks are posted to the current node.
And there is no data communication operation between the current node and the others in the subsequent computation process.

\textbf{(2) $ClusterScheduler$ for $T_{NS}$ tasks.}

\begin{figure}[!ht]
\setlength{\abovecaptionskip}{0pt}
\setlength{\belowcaptionskip}{0pt}
\centering
\includegraphics[width=3.5in]{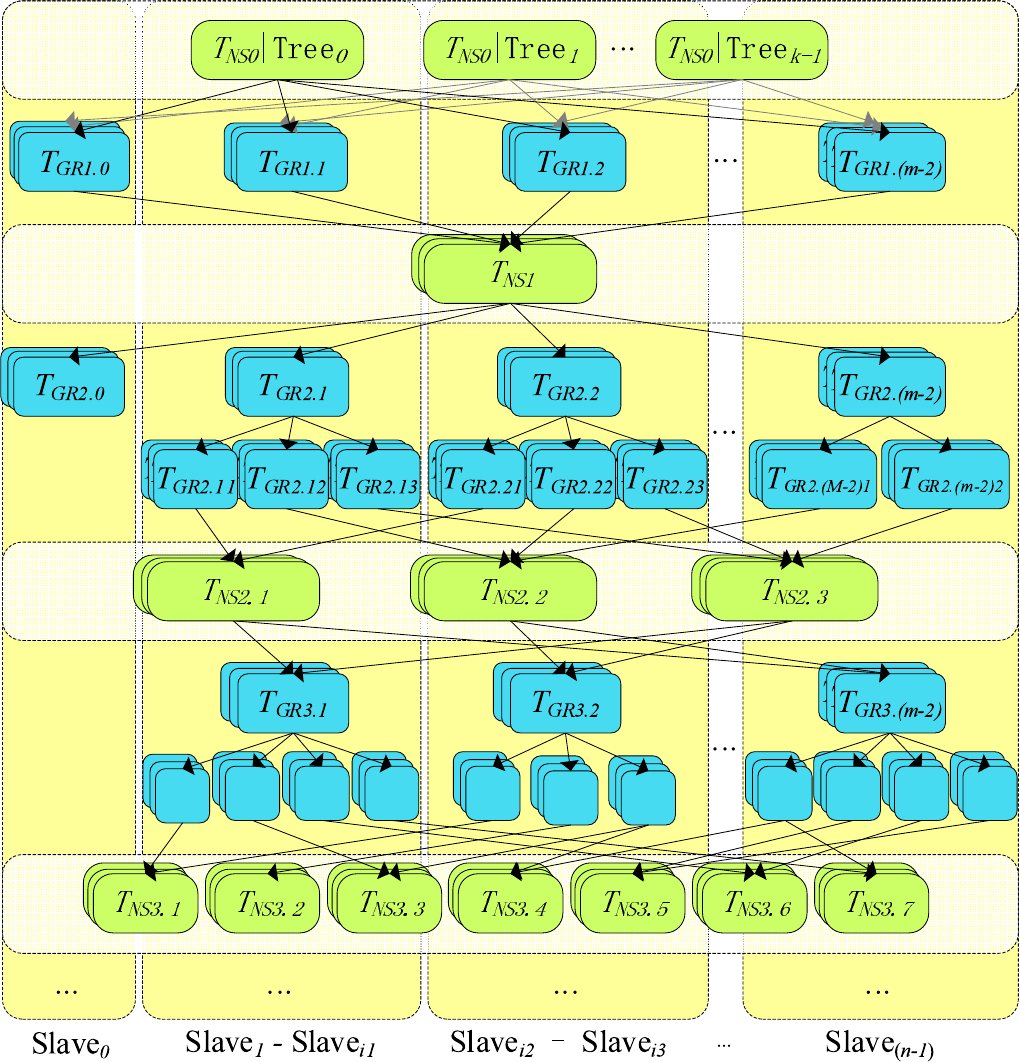}
\caption{Task-parallel scheduling based on the DAG in Fig. \ref{fig6}}
\label{fig7}
\end{figure}

The $ClusterScheduler$ module monitors the execution situation of the computing resources and tasks in the whole Spark cluster and allocates tasks to suitable workers.
As mentioned above, $T_{NS}$ tasks are used to collect the results of the corresponding $T_{GR}$ tasks and split the decision tree nodes.
$T_{NS}$ tasks are independent of all feature subsets and can be scheduled and allocated in the whole Spark cluster.
In addition, these $T_{NS}$ tasks rely on the results of the corresponding $T_{GR}$ tasks, therefore, there is a wait and synchronization restraint for these tasks.
Therefore, we invoke the $ClusterScheduler$ to perform $T_{NS}$ tasks.
We set the locality property value of each $T_{NS}$ as $ANY$ and submit to a $ClusterScheduler$ module.
The task-parallel scheduling scheme for $T_{NS}$ tasks is described in Algorithm \ref{alg42}.
A diagram of task-parallel scheduling for the tasks in the above DAG is shown in Fig. \ref{fig7}.

\begin{algorithm}
\caption{Task-parallel scheduling for $T_{NS}$ tasks}
\label{alg42}
\begin{algorithmic}[1]
\vspace{.2cm}
{\small
\REQUIRE ~\\
     $TS_{NS}$: a task set of all $T_{NS}$ submitted by $DAGScheduler$.
\ENSURE ~\\
     $ER_{TS}$: the execution results of $TS_{NS}$.
\STATE create $manager$ $\leftarrow$ new TaskSetManager($TS_{NS}$);
\STATE append to taskset manager $activeTSQueue$ $\leftarrow$ $manager$;
\IF { hasReceivedTask == false}
\STATE create $starvationTimer$ $\leftarrow$ scheduleAtFixedRate(new TimerTask);
\STATE rank the priority of $TS2$ $\leftarrow$ $activeTSQueue$.FIFO();
\FOR {each task $T_{i}$ in $TS2$}
\STATE get available worker $executor_{a}$ from $workers$;
\STATE $ER_{TS}$ $\leftarrow$ $executor_{a}$.launchTask($T_{i}$.taskid);
\ENDFOR
\ENDIF
\RETURN $ER_{TS}$.
}
\end{algorithmic}
\end{algorithm}

\subsection{Parallel Optimization Method Analysis}
We discuss our hybrid parallel optimization method from 5 aspects as follows.
In comparison with Spark-MLRF and other parallel methods of RF, our hybrid parallel optimization approach of PRF achieves advantages in terms of performance, workload balance, and scalability.

\subsubsection{Computational Complexity Analysis}
As discussed in Section 3.4, the total computational complexity of the improved PRF algorithm with dimension-reduction is $O(k(MN+mN \log N))$.
After the parallelization of the PRF algorithm on Spark, $M$ features of training dataset are calculated in parallel in the process of dimension-reduction, and $k$ trees are trained concurrently.
Therefore, the theoretical computational complexity of the PRF algorithm is $O(\frac{k(MN+mN \log N)}{k \times M}) \approx O(N(\log N + 1))$.

\subsubsection{Data Volume Analysis}
Taking advantage of the data-multiplexing method, the training dataset is reused effectively.
Assume that the volume of the original dataset is $(N \times M)$ and the RF model's scale is $k$, the volumes of the sampled training dataset in the original RF and Spark-MLRF are both $(N \times M \times k)$.
In our PRF, the volume of the sampled training dataset is $(N \times 2 \times (M-1)) \approx (2NM)$.
Moreover, the increase of the scale of PRF does not lead to changes in the data size and storage location.
Therefore, compared with the sampling method of the original RF and Spark-MLRF, the data-parallel method of our PRF decreases the total volume of the training dataset for PRF.

\subsubsection{Data Communication Analysis}
In PRF, there exist data communication operations in the process of data allocation and the training process.
Assume that there are $n$ slaves in a Spark cluster, and the data volume of the sampled training dataset is $(2NM)$.
In the process of data allocation, the average data communication cost is $(\frac{2MN}{n})$.
In the process of the PRF model training, if a feature subset is allocated to several computer nodes, local data communication operations of the subsequent computing tasks occur among these nodes.
If one or more feature subsets are allocated to one computer node, there is no data communication operation among different nodes in the subsequent computation process.
Generally, there is a small amount of data communication cost for the intermediate results in each stage of the decision tree's training process.
The vertical data-partitioning and static data allocation method mitigates the amount of data communication in the distributed environment and overcomes the performance bottleneck of the traditional parallel method.

\subsubsection{Resource and Workload Balance Analysis}
\label{WorkloadBalanceAnalysis}
From the view point of the entire training process of PRF in the whole Spark cluster, our hybrid parallel optimization approach achieves a better storage and workload balance than other algorithms.
One reason is that because the different volumes of feature subsets might lead to different workloads of the $T_{GR}$ tasks for each feature variable, we allocate the feature subsets to the Spark cluster according to its volume.
A feature subset with a large volume is allocated to multiple slave nodes. And the corresponding $T_{GR}$ tasks are scheduled among these nodes in parallel.
A feature subsets with a small volume are allocated to one slave node. And the corresponding $T_{GR}$ tasks are scheduled on the current node.

A second reason is that with the tree nodes' splitting, the slave nodes where the split variables' feature subsets are located will revert to an idle status.
From the view point of the entire training process of PRF, profit from the data-multiplexing method of PRF, each feature subset is shared and reused by all decision trees, and it might be split for different tree nodes in different trees.
That is, although a feature subset is split and useless to a decision tree, it is still useful to other trees.
Therefore, our PRF not only does not lead to the problem of waste of resources and workload imbalance, but also makes full use of the data resources and achieves an overall workload balance.

\subsubsection{Algorithm Scalability Analysis}
We discuss the stability and scalability of our PRF algorithm from 3 perspectives.
(1) The data-multiplexing method of PRF makes the training dataset be reused effectively.
When the scale of PRF expands, namely, the number of decision trees increases, the data size and the storage location of the feature subsets need not change.
It only results in an increase in computing tasks for new decision trees and a small amount of data communication cost for the intermediate results of these tasks.
(2) When the Spark cluster's scale expands, only a few feature subsets with a high storage load are migrated to the new computer nodes to achieve storage load and workload balance.
(3) When the scale of the training dataset increases, it is only necessary to split feature subsets from the new data in the same vertical data-partitioning way, and append each new subset to the corresponding original one.
Therefore, we can draw the conclusion that our PRF algorithm with the hybrid parallel optimization method achieves good stability and scalability.

\section{Experiments}
\subsection{Experiment Setup}
All the experiments are performed on a Spark cloud platform, which is built of one master node and 100 slave nodes.
Each node executes in Ubuntu 12.04.4 and has one Pentium (R) Dual-Core 3.20GHz CPU and 8GB memory.
All nodes are connected by a high-speed Gigabit network and are configured with Hadoop 2.5.0 and Spark 1.1.0.
The algorithm is implemented in Scala 2.10.4.
Two groups of datasets with large scale and high dimensionality are used in the experiments.
One is from the UCI machine learning repository \cite{ex33}, as shown in Table \ref{table51}.
Another is from a actual medical project, as shown in Table \ref{table52}.

\begin{table}[!ht]
\renewcommand{\arraystretch}{1.3}
\setlength{\abovecaptionskip}{0pt}
\setlength{\belowcaptionskip}{0pt}
\caption{Datasets from the UCI machine learning repository}
\label{table51}
\tabcolsep1pt
\begin{tabular*}{3.5in}{p{0.9in}l p{0.5in}c p{0.5in}c p{0.4in}c p{0.6in}c p{0.6in}c}
\hline
 Datasets & Instances & Features & Classes & Data Size & Data Size\\
&&&&(Original)&(Maximum)\\
\hline
URL Reputation (URL)  &	2396130	& 3231961 &	5 &	2.1GB  &	1.0TB  \\
You Tube Video Games (Games) &	120000 & 1000000 & 14 &	25.1GB & 2.0TB\\
Bag of Words (Words)	  & 8000000 & 100000  &	24 & 15.8GB &	1.3TB \\
Gas sensor arrays (Gas)& 1800000 & 1950000 & 15 & 50.2GB & 2.0TB \\
\hline
\end{tabular*}
\end{table}

\begin{table}[!ht]
\setlength{\abovecaptionskip}{0pt}
\setlength{\belowcaptionskip}{0pt}
\tabcolsep1pt
\renewcommand{\arraystretch}{1.3}
\caption{Datasets from a medical project}
\label{table52}
\begin{tabular*}{3.5in}{p{0.9in}l p{0.5in}c p{0.5in}c p{0.4in}c p{0.6in}c p{0.6in}c}
\hline
 Datasets & Instances & Features & Classes & Data size & Data size\\
&&&&(Original)&(Maximum)\\
\hline
Patient    & 279877	 & 25652 & 18 &	3.8GB  & 1.5TB\\
Outpatient & 3657789 & 47562 & 9  &	10.6GB & 1.0TB\\
Medicine   & 7502058 & 52460 & 12 &	20.4GB & 2.0TB\\
Cancer	   & 3568000 & 46532 & 21 &	5.8GB  & 2.0TB\\
\hline
\end{tabular*}
\end{table}

In Table \ref{table51} and Table \ref{table52}, $Data size (Original)$ refers to the original size of the data from the UCI and the project, and $Data size (Maximum)$ refers to the peak size of data sampled by all of the comparison algorithms.

In the Spark platform, the training data not be loaded into the memory as a whole.
Spark can be used to process datasets that are greater than the total cluster memory capacity.
RDD objects in a single executor process are accessed by an iteration, and the data are buffered or thrown away after the processing.
The cost of memory is very small when there is no requirement of caching the results of the RDD objects.
In this case, the results of the iterations are retained in a memory pool by the cache manager.
When the data in the memory are not applicable, they will be saved on disk.
In this case, part of the data can be kept in the memory and the rest is stored in the disk.
Therefore, the training data with the peak size of 2.0TB can be executed on Spark.

\subsection{Classification Accuracy}
We evaluate the classification accuracy of PRF by comparison with RF, DRF, and Spark-MLRF.

\subsubsection{Classification Accuracy for Different Tree Scales}
\begin{figure}[!ht]
\setlength{\abovecaptionskip}{0pt}
\setlength{\belowcaptionskip}{0pt}
\centering
\includegraphics[width=3.5in]{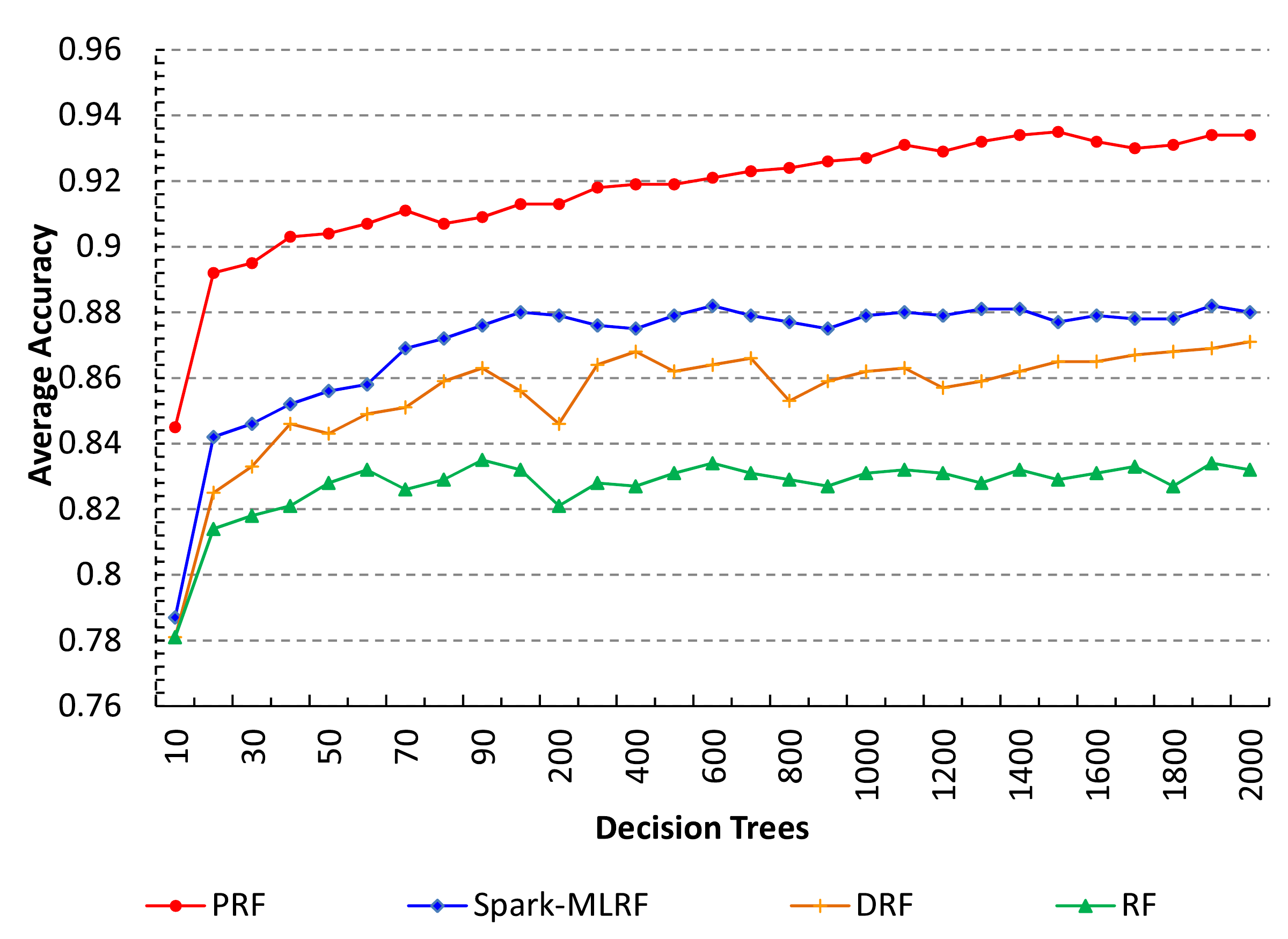}
\caption{Average classification accuracy for different tree scales}
\label{chart1}
\end{figure}

To illustrate the classification accuracy of PRF, experiments are performed for the RF, DRF \cite{ex18}, Spark-MLRF, and PRF algorithms.
The datasets are outlined in Table \ref{table51} and Table \ref{table52}.
Each case involves different scales of the decision tree.
The experimental results are presented in Fig. \ref{chart1}.

Fig. \ref{chart1} shows that the average classification accuracies of all of the comparative algorithms are not high when the number of decision trees is equal to 10.
As the number of decision trees increases, the average classification accuracies of these algorithms increase gradually and have a tendency toward a convergence.
The classification accuracy of PRF is higher than that of RF by 8.9\%, on average, and 10.6\% higher in the best case when the number of decision trees is equal to 1500.
It is higher than that of DRF by 6.1\%, on average, and 7.3\% higher in the best case when the number of decision trees is equal to 1300.
The classification accuracy of PRF is higher than that of Spark-MLRF by 4.6\% on average, and 5.8\% in the best case when the number of decision trees is equal to 1500.
Therefore, compared with RF, DRF, and Spark-MLRF, PRF improves the classification accuracy significantly.

\subsubsection{Classification Accuracy for Different Data Sizes}
\begin{figure}[!ht]
\setlength{\abovecaptionskip}{0pt}
\setlength{\belowcaptionskip}{0pt}
\centering
\includegraphics[width=3.5in]{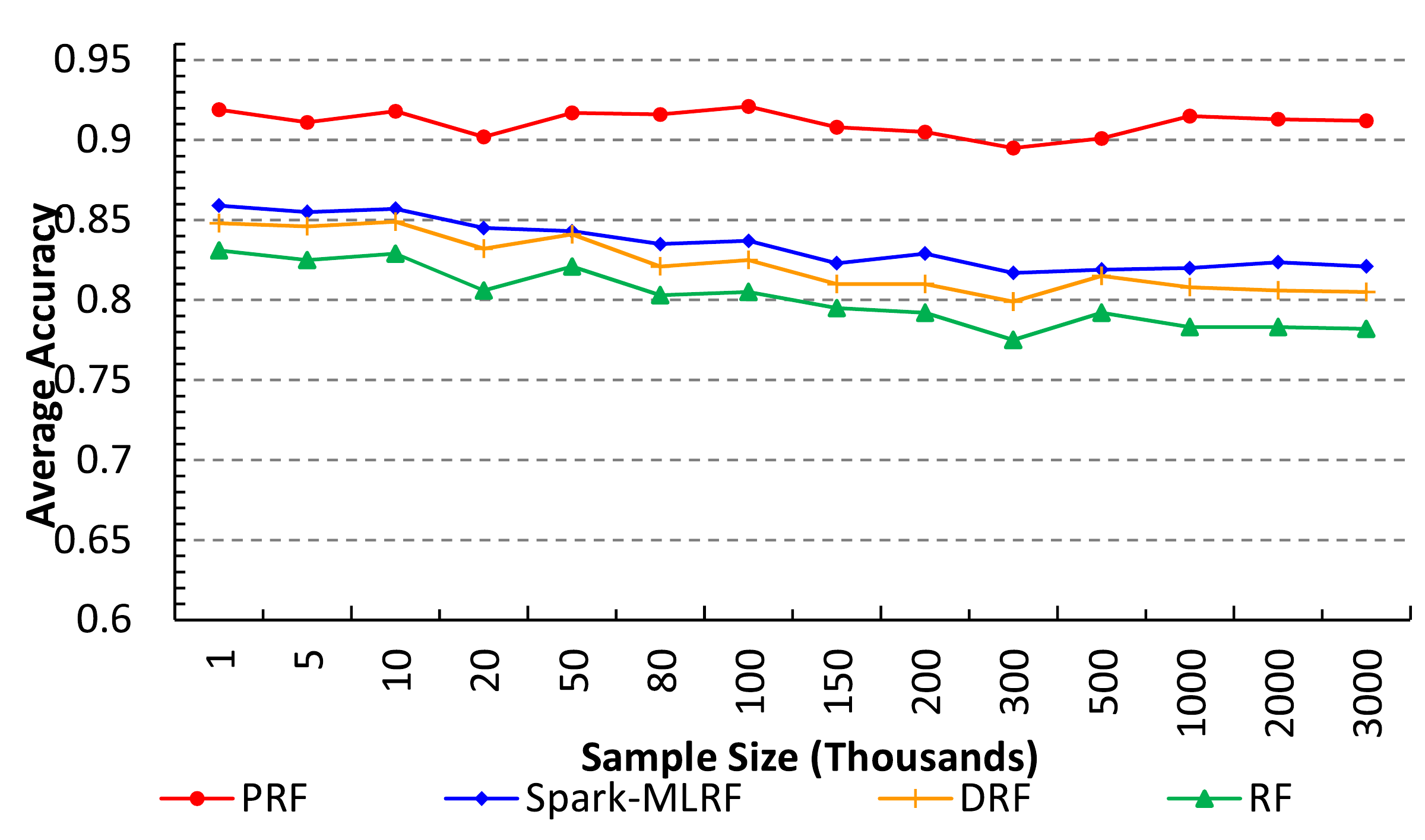}
\caption{Average classification accuracy for different data sizes}
\label{chart2}
\end{figure}

Experiments are performed to compare the classification accuracy of PRF with the RF, DRF, and Spark-MLRF algorithms.
Datasets from the project described in Table \ref{table52} are used in the experiments.
The experimental results are presented in Fig. \ref{chart2}.

The classification accuracies of PRF in all of the cases are greater than that of RF, DRF, and Spark-MLRF obviously for each scale of data.
The classification accuracy of PRF is greater than that of DRF by 8.6\%, on average, and 10.7\% higher in the best case when the number of samples is equal to 3,000,000.
The classification accuracy of PRF is greater than that of Spark-MLRF by 8.1\%, on average, and 11.3\% higher in the best case when the number of samples is equal to 3,000,000.
For Spark-MLRF, because of the method of sampling for each partition of the dataset, as the size of the dataset increases, the ratio of the random selection of the dataset increases, and the accuracy of Spark-MLRF decreases inevitably.
Therefore, compared with RF, DRF, and Spark-MLRF, PRF improves the classification accuracy significantly for different scales of datasets.

\subsubsection{OOB Error Rate for Different Tree Scales}
\begin{figure}[!ht]
\setlength{\abovecaptionskip}{0pt}
\setlength{\belowcaptionskip}{0pt}
\centering
\includegraphics[width=3.5in]{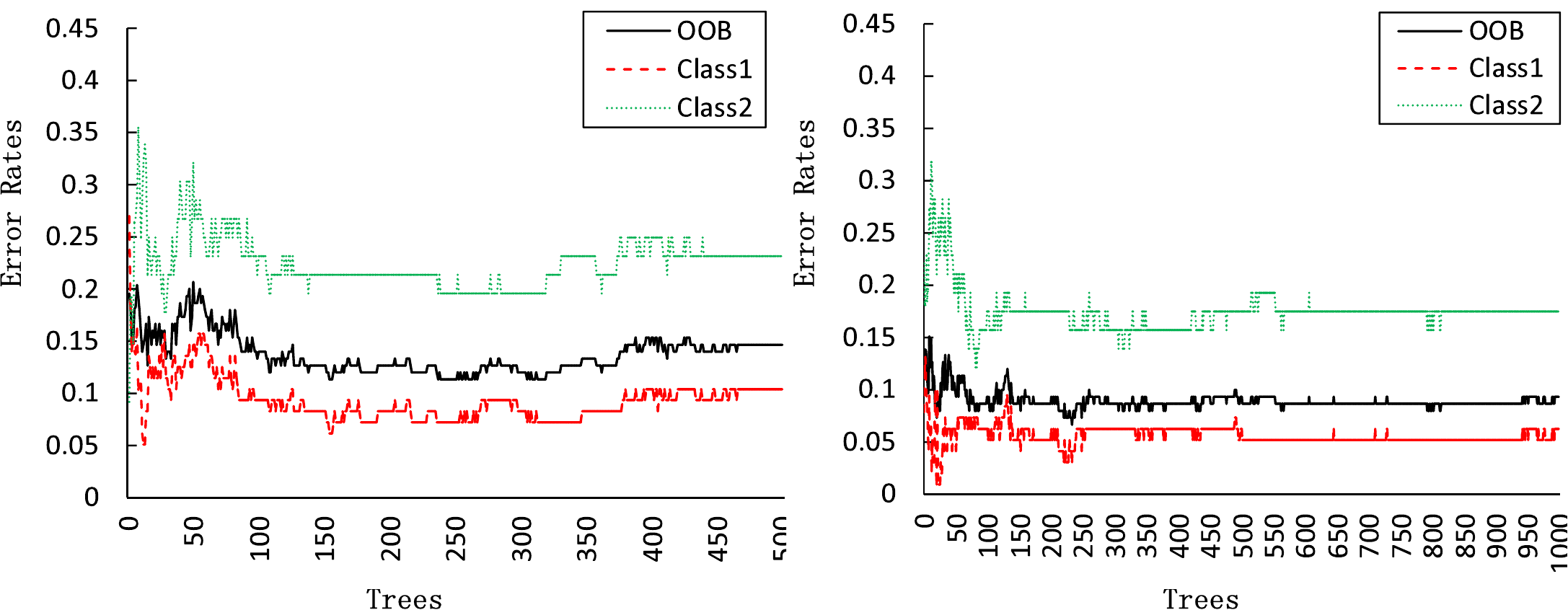}
\caption{OOB error rates of PRF for different tree scales}
\label{chart3}
\end{figure}

We observe the classification error rate of PRF under different conditions.
In each condition, the $Patient$ dataset is chosen, and two scales (500 and 1000) of decision trees are constructed.
The experimental results are presented in Fig. \ref{chart3} and Table \ref{table54}.

When the number of decision trees of PRF increases, the OOB error rate in each case declines gradually and tends to a convergence condition.
The average OOB error rate of PRF is 0.138 when the number of decision trees is equal to 500, and it is 0.089 when the number of decision trees is equal to 1000.

\begin{table}[!ht]
\setlength{\abovecaptionskip}{0pt}
\setlength{\belowcaptionskip}{0pt}
\renewcommand{\arraystretch}{1.3}
\caption{OOB error rates of PRF for different tree scales}
\label{table54}
\begin{tabular*}{3.5in}{c c c c c c c}
\hline
	 &\multicolumn{3}{c}{Tree=500}&\multicolumn{3}{c}{Tree=1000}	 \\
Rate & OOB	 & Class1 & Class2 & OOB   & Class1 & Class2\\
\hline
max	 & 0.207 & 0.270  &	0.354  & 0.151 & 0.132  & 0.318 \\
min	 & 0.113 & 0.051  &	0.092  & 0.067 & 0.010  & 0.121 \\
mean & 0.138 & 0.094  &	0.225  & 0.089 & 0.056  & 0.175 \\
\hline
\end{tabular*}
\end{table}

\subsection{Performance Evaluation}
Various experiments are constructed to evaluate the performance of PRF by comparison with the RF and Spark-MLRF algorithms in terms of the execution time, speedup, data volume, and data communication cost.

\subsubsection{Average Execution Time for Different Datasets}
\begin{figure}[!ht]
\setlength{\abovecaptionskip}{0pt}
\setlength{\belowcaptionskip}{0pt}
\centering
\includegraphics[width=3.5in]{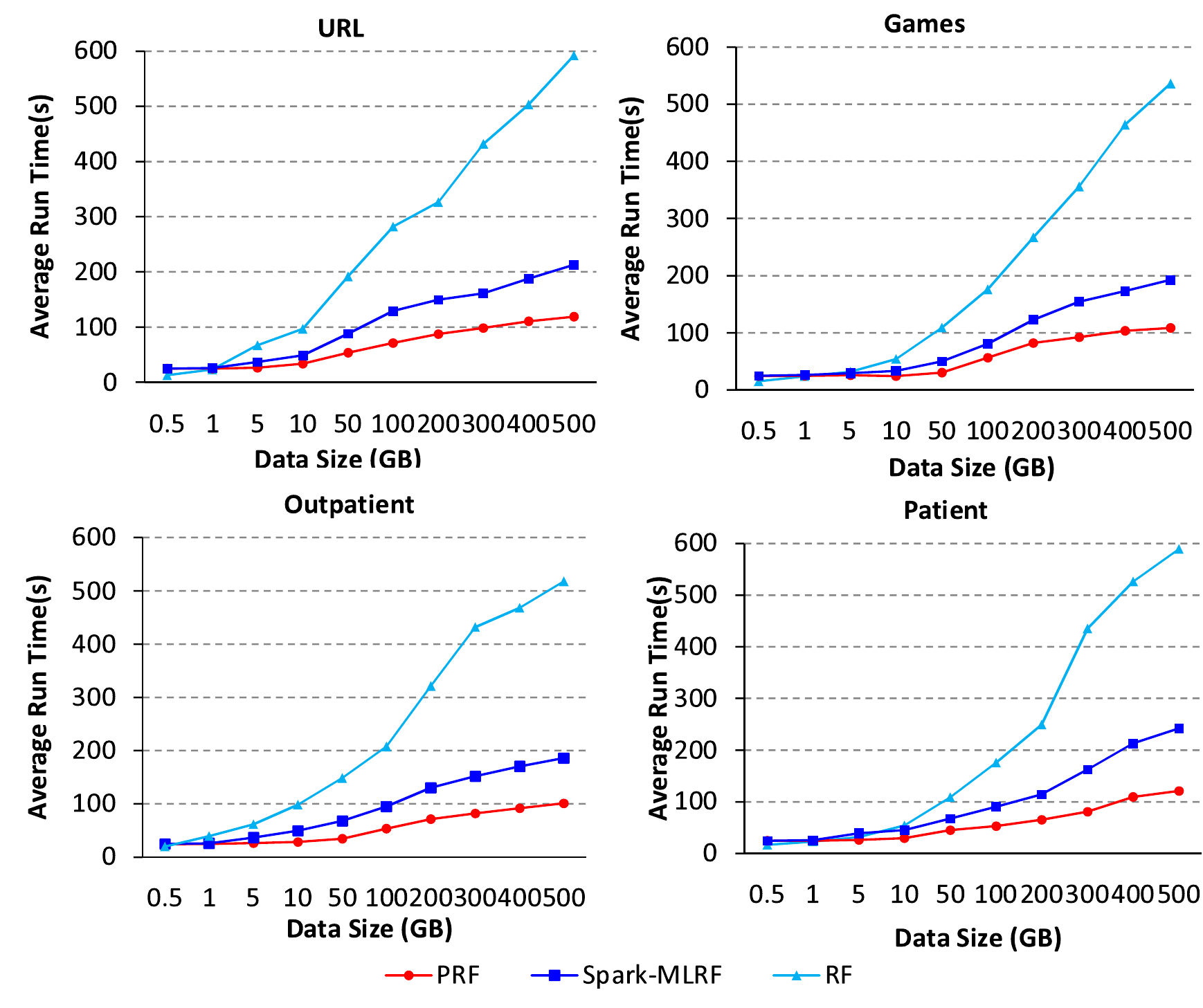}
\caption{Average execution time of the algorithms for different datasets}
\label{chart4}
\end{figure}

Experiments are performed to compare the performance of PRF with that of RF and Spark-MLRF.
Four groups of training datasets are used in the experiments, such as $URL$, $Games$, $Outpatient$, and $Patient$.
In these experiments, the number of decision trees in each algorithm is both 500, and the number of Spark slaves is 10.
The experimental results are presented in Fig. \ref{chart4}.

When the data size is small (e.g., less than 1.0GB), the execution times of PRF and Spark-MLRF are higher than that of RF.
The reason is that there is a fixed time required to submit the algorithms to the Spark cluster and configure the programs.
When the data size is greater than 1.0GB, the average execution times of PRF and Spark-MLRF are less than that of RF in the four cases.
For example,  in the $Outpatient$ case, when the data size grows from 1.0 to 500.0GB, the average execution time of RF increases from 19.9 to 517.8 seconds, while that of Spark-MLRF increases from 24.8 to 186.2 seconds, and that of PRF increases from 23.5 to 101.3 seconds.
Hence, our PRF algorithm achieves a faster processing speed than RF and Spark-MLRF.
When the data size increases, the benefit is more noticeable.
Taking advantage of the hybrid parallel optimization, PRF achieves significant strengths over Spark-MLRF and RF in terms of performance.

\subsubsection{Average Execution Time for Different Cluster Scales}
\begin{figure}[!ht]
\setlength{\abovecaptionskip}{0pt}
\setlength{\belowcaptionskip}{0pt}
\centering
\includegraphics[width=3.2in]{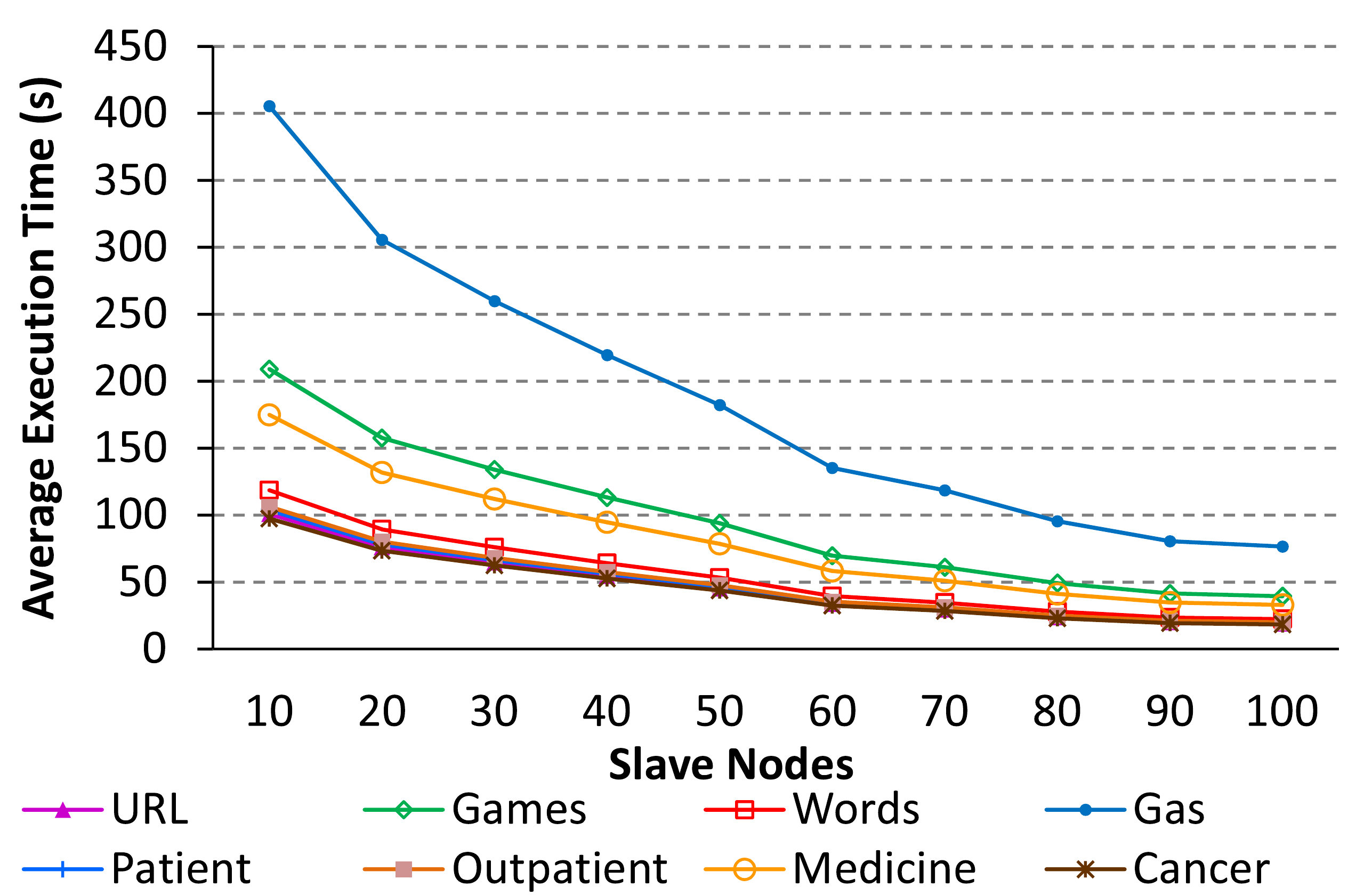}
\caption{Average execution time of PRF for different cluster scales}
\label{chart5}
\end{figure}

In this section, the performance of PRF on the Spark platform for different scales of slave nodes is considered.
The number of slave nodes is gradually increased from 10 to 100, and the experiment results are presented in Fig. \ref{chart5}.

In Fig. \ref{chart5}, because of the different data sizes and contents of the training data, the execution times of PRF in each case are different.
When the number of slave nodes increases from 10 to 50, the average execution times of PRF in all cases obviously decrease.
For example, the average execution time of PRF decreases from 405.4 to 182.6 seconds in the $Gas$ case and from 174.8 to 78.3 seconds in the $Medicine$ case.
By comparison, the average execution times of PRF in the other cases decrease less obviously when the number of slave nodes increases from 50 to 100.
For example, the average execution time of PRF decreases from 182.4 to 76.0 seconds in the $Gas$ case and from 78.3 to 33.0 seconds in the $Medicine$ case.
This is because when the number of the Spark slaves greater than that of training dataset's feature variables, each feature subset might be allocated to multiple slaves.
In such a case, there are more data communication operations among these slaves than before, which leads to more execution time of PRF.

\subsubsection{Speedup of PRF in Different Environments}
Experiments in a stand-alone environment and a Spark cloud platform are performed to evaluate the speedup of PRF.
Because of the different volume of training datasets, the execution times of PRF are not in the same range in different cases.
To observe the comparison of the execution time intuitively, a normalization of execution time is taken.
Let $T_{(i, sa)}$ be the execution time of PRF for dataset $S_{i}$ in the stand-alone environment, and first normalized to 1.
The execution time of PRF on Spark is normalized as described in Eq. (\ref{equ16}):

\begin{equation}
\label{equ16}
T_{i}^{'}=\left\{
\begin{array}{lcl}
\frac{T_{(i, sa)}}{T_{(i, sa)}} =1&   &Stand-alone,\\
\frac{T_{(i,Spark)}}{T_{(i,sa)}} &   &Spark.
\end{array} \right.
\end{equation}

The speedup of PRF on Spark for $S_{i}$ is defined in Eq. (\ref{equ17}):

\begin{equation}
\label{equ17}
Speedup_{(i,Spark)}= \frac{T_{(i,Spark)}^{'}}{T_{(i,sa)}^{'}}.
\end{equation}

\begin{figure}[!ht]
\setlength{\abovecaptionskip}{0pt}
\setlength{\belowcaptionskip}{0pt}
\centering
\includegraphics[width=3.5in]{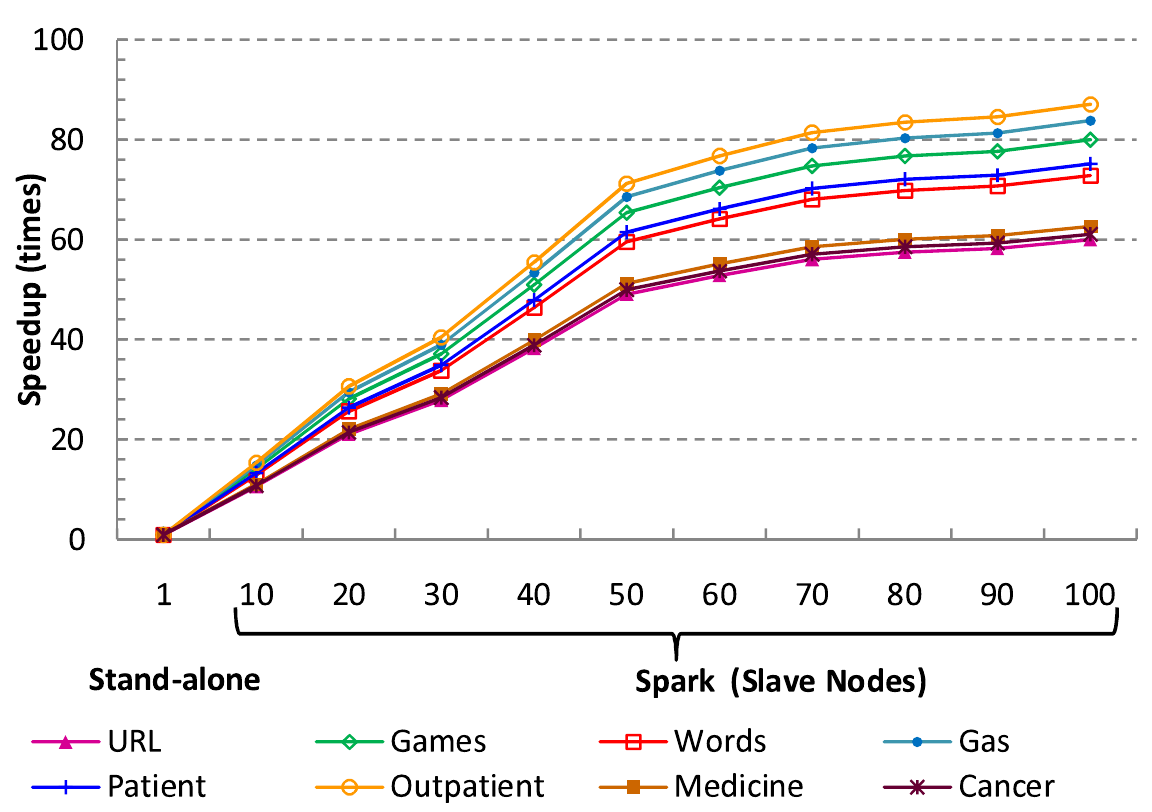}
\caption{Speedup of PRF in different environments }
\label{chart6}
\end{figure}

The results of the comparative analyses are presented in Fig. \ref{chart6}.
Taking benefits of the parallel algorithm and cloud environment, the speedup of PRF on Spark tends to increase in each experiment with the number of slave nodes.
When the number of slave nodes is equal to 100, the speedup factor of PRF in all cases is in the range of 60.0 - 87.3, which is less than the theoretical value (100).
Because there exists data communication time in a distributed environment and a fixed time for the application submission and configuration, it is understandable that the whole speedup of PRF is less than the theoretical value.
Due to the different data volumes and contents, the speedup of PRF in each case is different.

When the number of slave nodes is less than 50, the speedup in each case shows a rapid growth trend.
For instance, compared with the stand-alone environment, the speedup factor of $Gas$ is 65.5 when the number of slave nodes is equal to 50, and the speedup factor of $Patient$ is 61.5.
However, the speedup in each case shows a slow growth trend when the number of slave nodes is greater than 50.
This is because there are more data allocation, task scheduling, and data communication operations required for PRF.

\subsubsection{Data Volume Analysis for Different RF Scales}
\begin{figure}[!ht]
\setlength{\abovecaptionskip}{0pt}
\setlength{\belowcaptionskip}{0pt}
\centering
\includegraphics[width=3.5in]{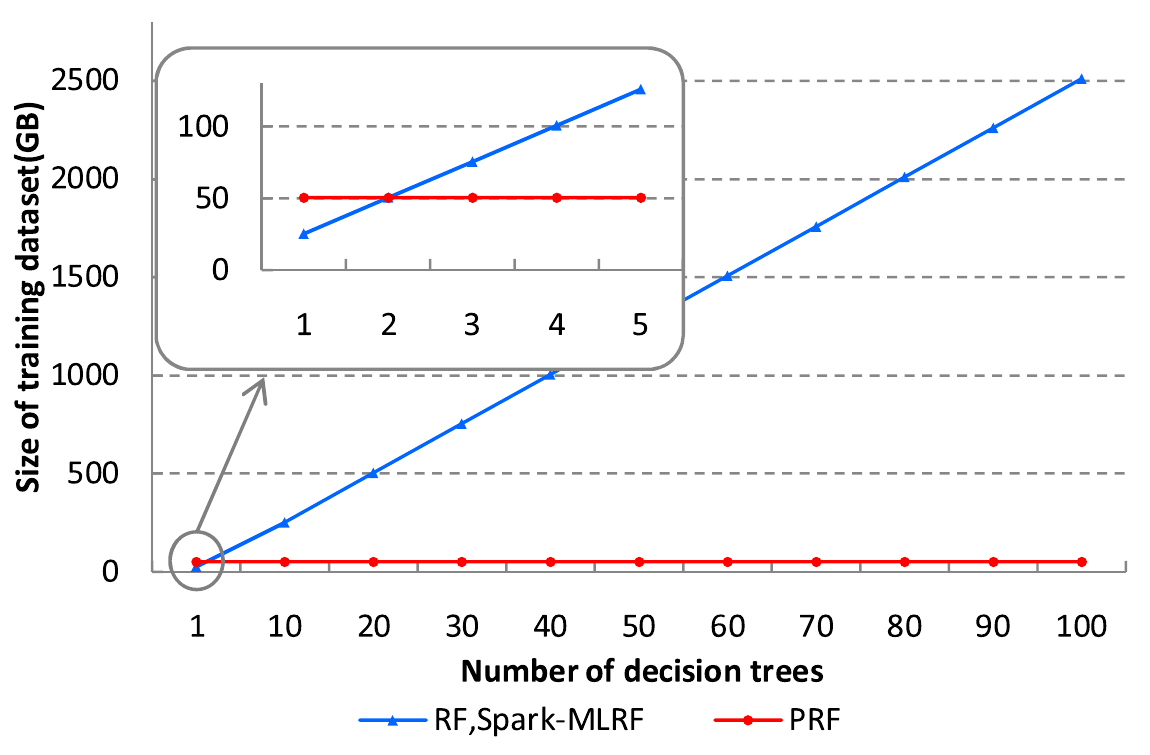}
\caption{Size of training dataset for different RF scales}
\label{chart7}
\end{figure}

We analyze the volume of the training data in PRF against RF and Spark-MLRF.
Taking the $Games$ case as an example, the volumes of the training data in the different RF scales are shown in Fig. \ref{chart7}.

In Fig. \ref{chart7}, due to the use of the same horizontal sampling method, the training data volumes of RF and Spark-MLRF both show a linear increasing trend with the increasing of the RF model scale.
Contrary, in PRF, the total volume of all training feature subsets is 2 times the size of the original training dataset.
Making use of the data-multiplexing approach of PRF, the training dataset is effectively reused.
When the number of decision trees is larger than 2, despite the expansion of RF scale, the volume of the training data will not increases any further.

\subsubsection{Data Communication Cost Analysis}
\begin{figure}[!ht]
\setlength{\abovecaptionskip}{0pt}
\setlength{\belowcaptionskip}{0pt}
\centering
\includegraphics[width=3.5in]{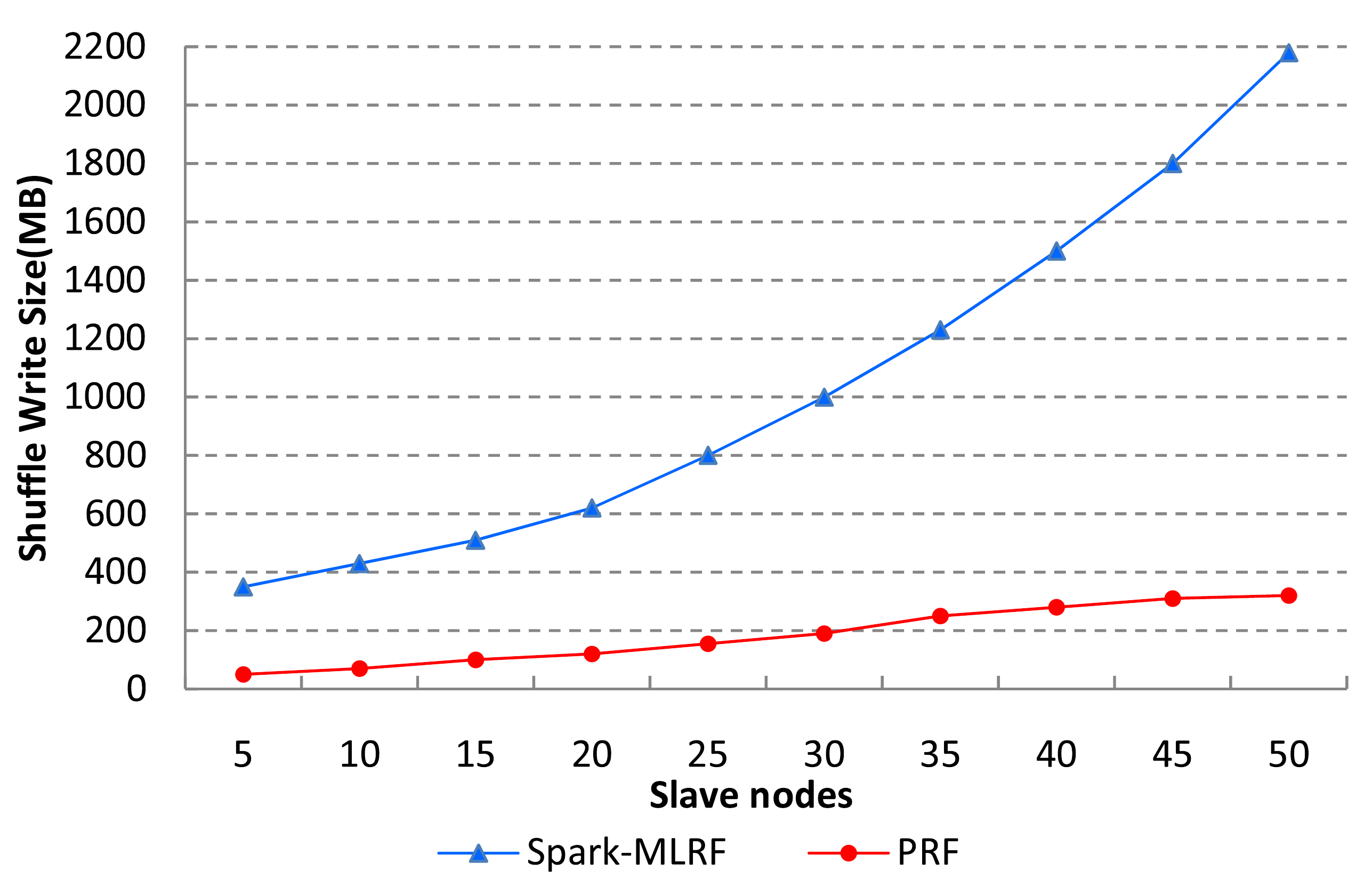}
\caption{Data communication costs of PRF and Spark-MLRF}
\label{chart8}
\end{figure}

Experiments are performed for different scales of the Spark cluster to compare the Data Communication Cost ($C_{DC}$) of PRF with that of Spark-MLRF.
The suffer-write size of slave nodes in the Spark cluster is monitored as the $C_{DC}$ of the algorithms.
Taking the $Patient$ case as an example, the results of the comparison of $C_{DC}$ are presented in Fig. \ref{chart8}.

From Fig. \ref{chart8}, it is clear that the $C_{DC}$ of PRF are less than that of Spark-MLRF in all cases, and the distinction is larger with increasing number of slave nodes.
Although Spark-MLRF also uses the data-parallel method, the horizontal partitioning method for training data makes the computing tasks have to frequent access data across different slaves.
As the number of slaves increases from 5 to 50, the $C_{DC}$ of Spark-MLRF increases from 350.0MB to 2180.0MB.
Different from Spark-MLRF, in PRF, the vertical data-partitioning and allocation method and the task scheduling method make the most of the computing tasks ($T_{GR}$) access data from the local slave, reducing the amount of data transmission in the distributed environment.
As the number of slaves increases from 5 to 50, the $C_{DC}$ of PRF increases from 50.0MB to 320.0MB, which is much lower than that of Spark-MLRF.
Therefore, PRF minimizes the $C_{DC}$ of RF in a distributed environment.
The expansion of the cluster's scale does not lead to an obviously increase in $C_{DC}$.
In conclusion, our PRF achieves a superiority and notable advantages over Spark-MLRF in terms of stability and scalability.

\section{Conclusions}
In this paper, a parallel random forest algorithm has been proposed for big data.
The accuracy of the PRF algorithm is optimized through dimension-reduction and the weighted vote approach.
Then, a hybrid parallel approach of PRF combining data-parallel and task-parallel optimization is performed and implemented on Apache Spark.
Taking advantage of the data-parallel optimization, the training dataset is reused and the volume of data is reduced significantly.
Benefitting from the task-parallel optimization, the data transmission cost is effectively reduced and the performance of the algorithm is obviously improved.
Experimental results indicate the superiority and notable strengths of PRF over the other algorithms in terms of classification accuracy, performance, and scalability.
For future work, we will focus on the incremental parallel random forest algorithm for data streams in cloud environment, and improve the data allocation and task scheduling mechanism for the algorithm on a distributed and parallel environment.

\section*{Acknowledgment}
The research was partially funded by
the Key Program of National Natural Science Foundation of China (Grant Nos. 61133005, 61432005),
the National Natural Science Foundation of China (Grant Nos. 61370095, 61472124, 61202109, 61472126,61672221),
the National Research Foundation of Qatar (NPRP, Grant Nos. 8-519-1-108),
and the Natural Science Foundation of Hunan Province of China (Grant Nos. 2015JJ4100, 2016JJ4002).

\ifCLASSOPTIONcaptionsoff
  \newpage
\fi

\bibliographystyle{IEEEtran}
\small
\bibliography{reference}

\begin{IEEEbiography}
[{\includegraphics[width=1in, height=1.25in, clip, keepaspectratio]{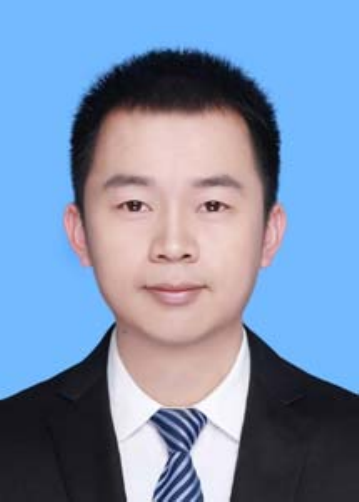}}]
{Jianguo Chen}  received the Ph.D. degree in College of Computer Science and Electronic Engineering at Hunan University, China.
He was a visiting Ph.D. student at the University of Illinois at Chicago from 2017 to 2018.
He is currently a postdoctoral in University of Toronto and Hunan University.
His major research areas include parallel computing, cloud computing, machine learning, data mining, bioinformatics and big data.
He has published research articles in international conference and journals of data-mining algorithms and parallel computing, such as
{\em IEEE TPDS}, IEEE/ACM TCBB, and {\em Information Sciences}.
\end{IEEEbiography}

\begin{IEEEbiography}
[{\includegraphics[width=1in, height=1.25in, clip, keepaspectratio]{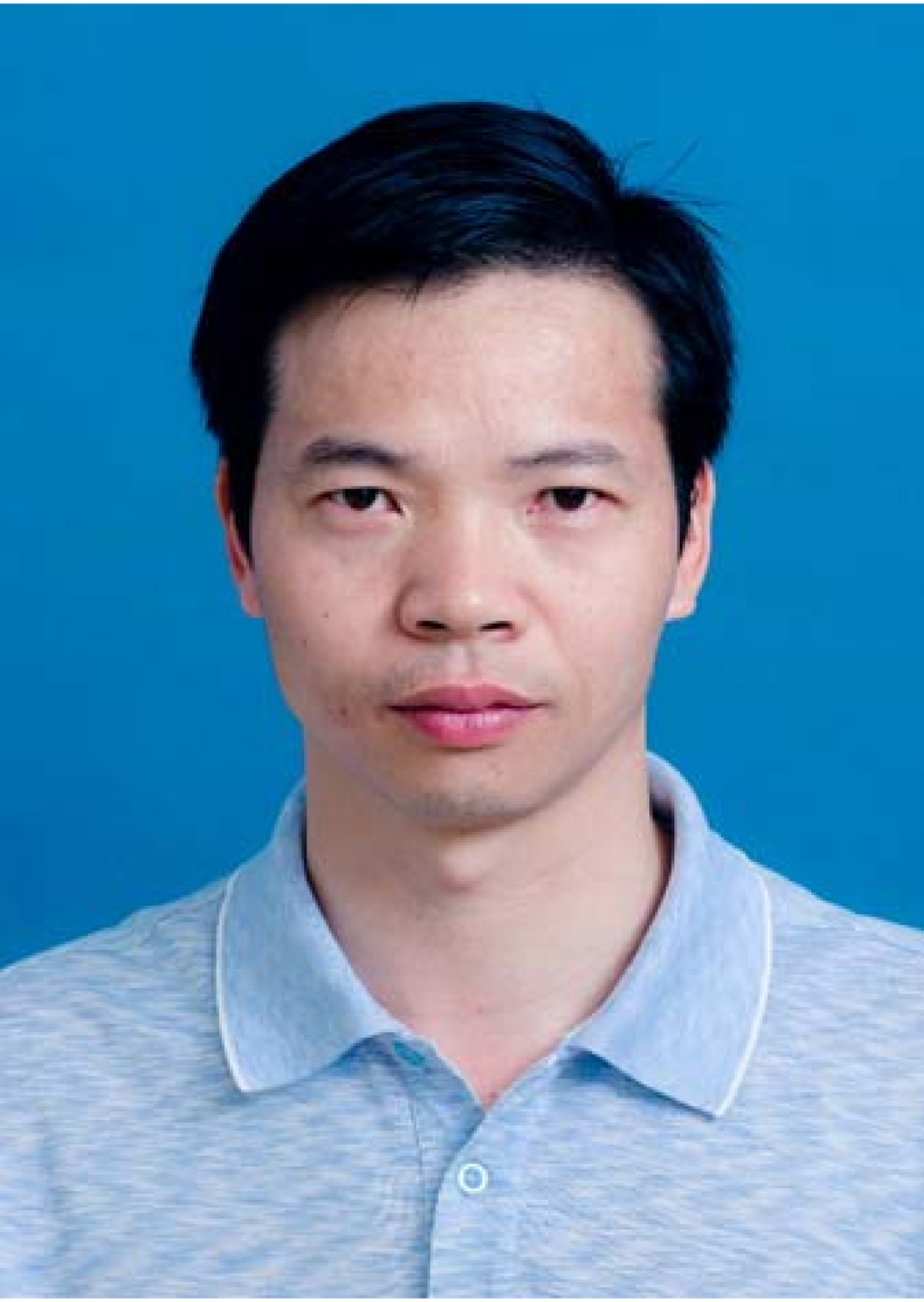}}]
{Kenli Li} received the Ph.D. degree in computer science from Huazhong University of Science and Technology, China, in 2003.
He was a visiting scholar at University of Illinois at Urbana-Champaign from 2004 to 2005.
He is currently a full professor of computer science and technology at Hunan University
and director of National Supercomputing Center in Changsha.
His major research areas include parallel computing, high-performance computing, grid and cloud computing.
He has published more than 180 research papers in international conferences and journals, such as
{\em IEEE-TC}, {\em IEEE-TPDS}, {\em IEEE-TSP}, {\em JPDC}, {\em ICPP}, {\em CCGrid}.
He is an outstanding member of CCF. He is a senior member of the IEEE and serves on the editorial board of {\em IEEE Transactions on Computers}.
\end{IEEEbiography}

\begin{IEEEbiography}
[{\includegraphics[width=1in, height=1.25in, clip, keepaspectratio]{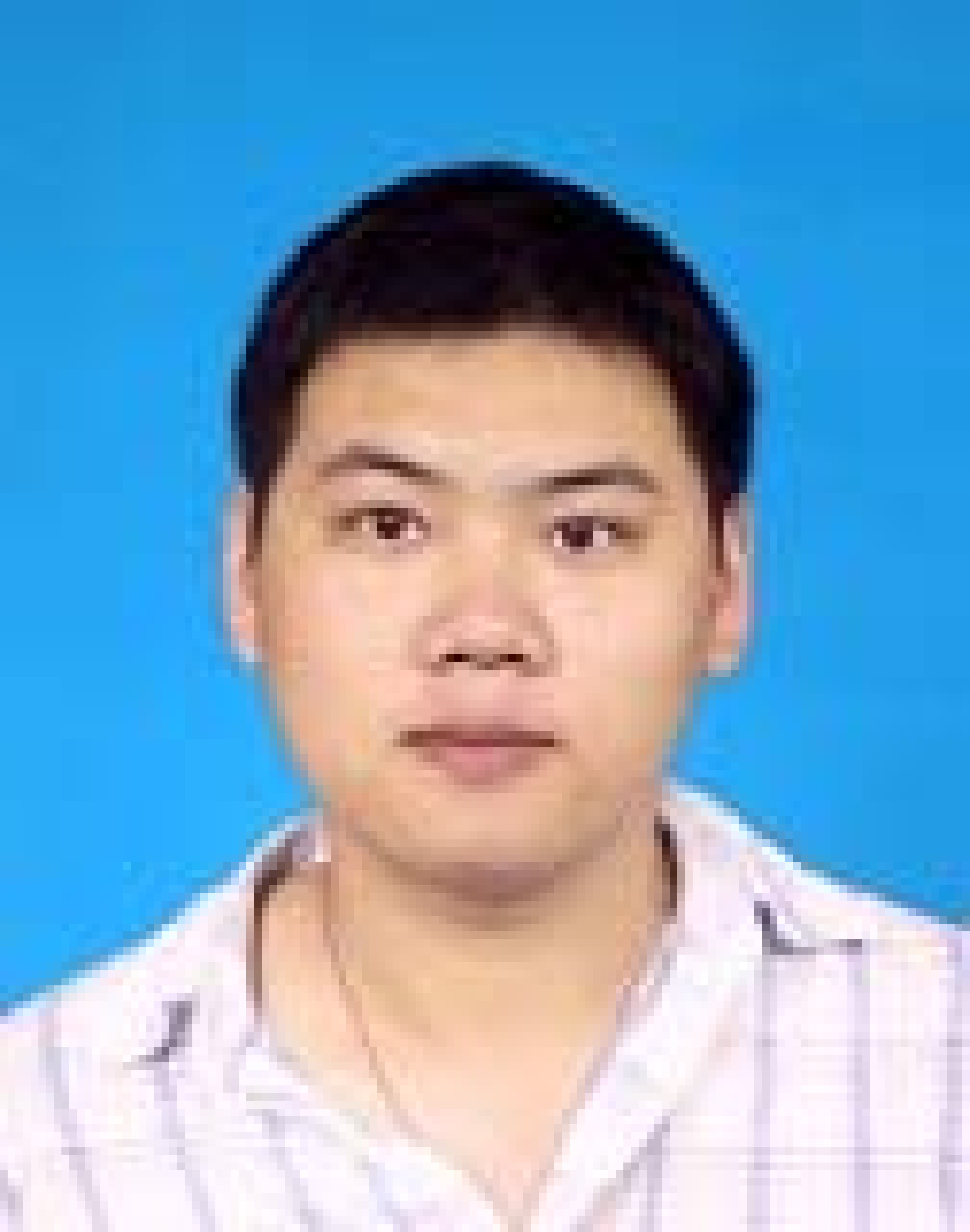}}]
{Zhuo Tang} received the Ph.D. in computer science from
Huazhong University of Science and Technology, China, in 2008. He is currently an associate professor of Computer Science and Technology at Hunan University. His research interests include security model, parallel algorithms, and resources scheduling for distributed computing systems, grid and cloud computing. He is a member of ACM and CCF.
\end{IEEEbiography}

\begin{IEEEbiography}
[{\includegraphics[width=1in, height=1.25in, clip, keepaspectratio]{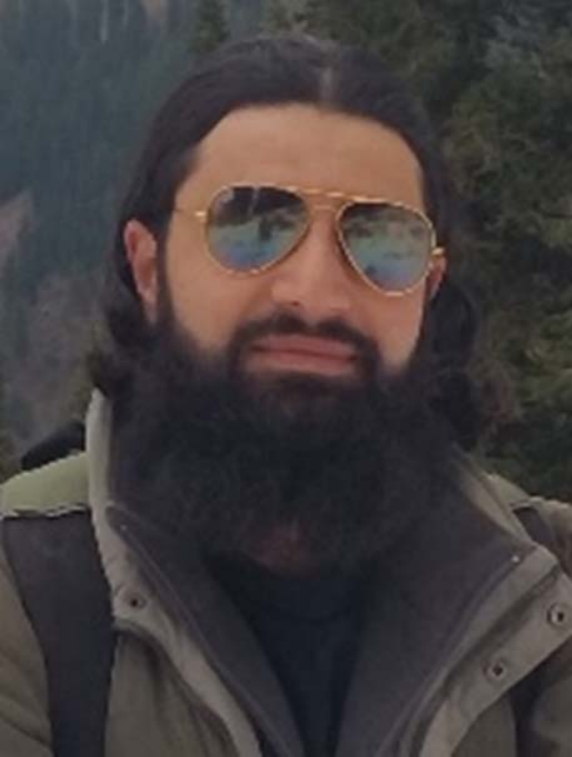}}]
{Kashif Bilal} received his PhD from North Dakota State University USA. He is currently a post-doctoral researcher at Qatar University, Qatar. His research interests include cloud computing, energy efficient high speed networks, and robustness.Kashif is awarded CoE Student Researcher of the year 2014 based on his research contributions during his doctoral studies at North Dakota State University.
\end{IEEEbiography}

\begin{IEEEbiography}
[{\includegraphics[width=1in, height=1.25in, clip, keepaspectratio]{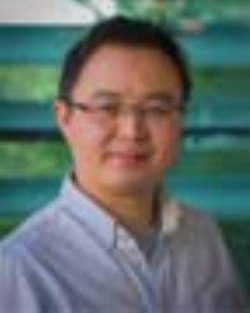}}]
{Shui Yu} is currently a Senior Lecturer of School of Information Technology, Deakin University. He is a member of Deakin University Academic Board (2015-2016), a Senior Member of IEEE, and a member of AAAS, the vice chair of Technical Subcommittee on Big Data Processing, Analytics, and Networking of IEEE Communication Society. He is currently serving the editorial boards of {\em IEEE TPDS}, {\em IEEE CST}, {\em IEEE Access}.
\end{IEEEbiography}

\begin{IEEEbiography}
[{\includegraphics[width=1in, height=1.25in, clip, keepaspectratio]{KashifBilal.pdf}}]
{Chuliang Weng} is a principal researcher at Huawei Shannon Lab. He received his Ph.D. from Shanghai Jiao Tong University in 2004, and from 2011 to 2012, he was a visiting scientist with the Department of Computer Science at Columbia University in the City of New York. His research interests include parallel and distributed systems, operating systems and virtualization, and storage systems. He is a member of the IEEE, ACM and CCF.
\end{IEEEbiography}

\begin{IEEEbiography}
[{\includegraphics[width=1in, height=1.25in, clip, keepaspectratio]{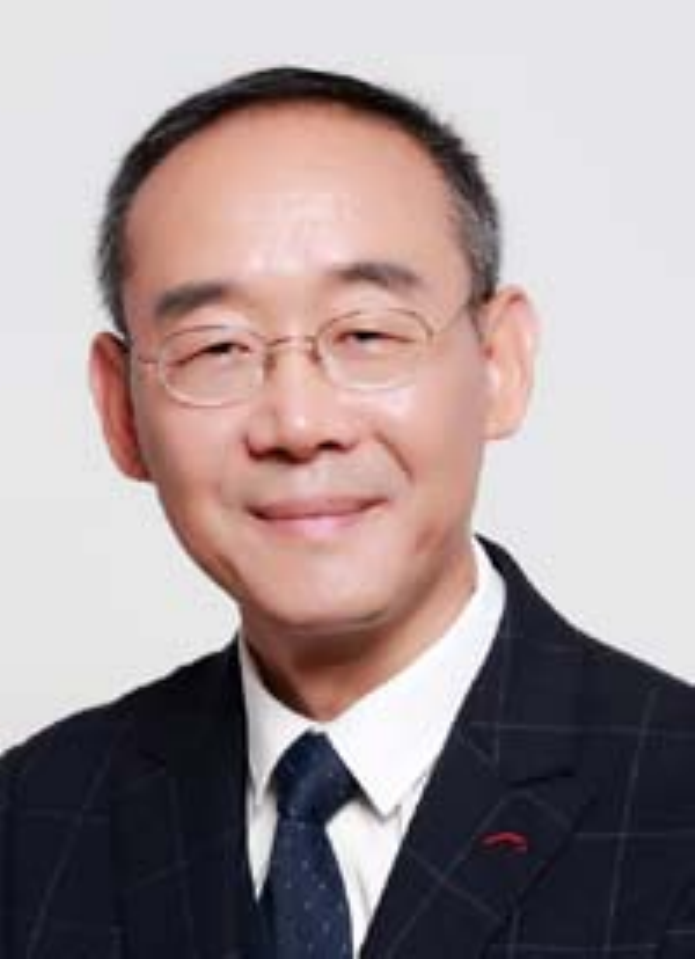}}]
{Keqin Li} is a SUNY Distinguished Professor of computer science in the State University of New York. His current research interests include parallel computing and high-performance computing, distributed computing, energy-efficient computing and communication, heterogeneous computing systems, cloud computing, big data computing, CPU-GPU hybrid and cooperative computing, multi-core computing, storage and file systems,
wireless communication networks, sensor networks, peer-to-peer file sharing systems, mobile computing, service computing, Internet of things and cyber-physical systems. He has
published over 590 journal articles, book chapters, and refereed conference papers, and
has received several best paper awards. He is currently serving or has served on the editorial boards of IEEE Transactions on Parallel and Distributed Systems, IEEE Transactions
on Computers, IEEE Transactions on Cloud Computing, IEEE Transactions on Services
Computing, and IEEE Transactions on Sustainable Computing. He is an IEEE Fellow.
\end{IEEEbiography}
\end{document}